\documentclass[prl,aps,amsmath,amssymb,superscriptaddress,longbibliography,notitlepage,twocolumn]{revtex4-1}
\usepackage{graphicx,multirow,outlines,ulem}
\usepackage{color}
\usepackage{hhline}
\usepackage{physics}
\usepackage{hyperref}

\usepackage{caption}
\usepackage{subcaption}
\usepackage{lipsum}

\usepackage{lipsum} 

\usepackage[sc,osf]{mathpazo}

\usepackage[section]{placeins}
\usepackage[mathscr]{eucal}
\newcommand{\huk}{\hookrightarrow}

\usepackage{tikz}
\usepackage{xcolor}

\newcommand{\bo}{\begin{outline}}
\newcommand{\eo}{\end{outline}}

\newcommand{\qed}{\nobreak \ifvmode \relax \else
      \ifdim\lastskip<1.5em \hskip-\lastskip
      \hskip1.5em plus0em minus0.5em \fi \nobreak
      \vrule height0.75em width0.5em depth0.25em\fi}
\begin{document}

\title{Witnessing  quantum chaos using observational entropy} 
\author{Sreeram PG}
\email{sreerampg@outlook.com}    
\affiliation{Department of Physics, Indian Institute of Science Education and Research, Pune 411008, India}
\author{Ranjan Modak}
\email{ranjan@iittp.ac.in}
\author{S. Aravinda}
\email{aravinda@iittp.ac.in}      
\affiliation{ Department of Physics, Indian Institute of Technology Tirupati, Tirupati,  517619, India} 
\begin{abstract}
We study observation entropy (OE) for the Quantum kicked top (QKT) model, whose classical
counterpart possesses different phases: regular, mixed, or chaotic, depending on the strength of
the kicking parameter. We show that OE grows logarithmically with coarse-graining length beyond a critical value in the regular phase, while OE growth is much faster in the chaotic regime. In the dynamics, we demonstrate that the short-time growth rate of OE acts as a
measure of the chaoticity in the system, and we compare our results with out-of-time-ordered correlators (OTOC). Moreover, we show that in the deep quantum regime, the results obtained from OE are much more robust compared to OTOC results. Finally, we also investigate the long-time behaviour of OE to distinguish between saddle-point scrambling and true chaos, where the former shows large persistent fluctuations compared to the latter.
\end{abstract}

\maketitle

\section{Introduction}
Classical chaos is one of the most significant discoveries in modern classical mechanics, and its emergence and applications in various research fields are paralleled with the development of classical computers. Around 1980, defining quantum chaos based on the correspondence principle was initiated, and the definition depends mainly on the chaotic behaviour of the classically limited system from its quantum counterpart~\cite{casati1979stochastic,shepelyansky1983some,berry1979quantum,dittrich1990long}. 
Random matrix theory has contributed significantly to defining and characterizing quantum chaos~\cite{haake1991quantum}. The emergence of quantum information theory and its applications in various fields of research, particularly in quantum many-body systems and quantum gravity accelerated the studies on the role of quantum chaos~\cite{hayden2007black,shenker2014black}. Such cross-field fertilization brought new tools and the application of quantum chaos to the forefront of research. The most important diagnostic tool is the Out-of-Time-Ordered Correlator (OTOC), from which the quantum Lyapunov exponent can be extracted~\cite{maldacena2016bound,larkin1969quasiclassical}. OTOC has been studied for various systems of interest, from simple systems to quantum many-body systems and continuous variable systems \cite{hashimoto2017out,larkin1969quasiclassical,swingle2018unscrambling,swingle2016measuring,sreeram2021out,maldacena2016bound,cotler2018out,chavez2019quantum,hashimoto2020exponential,garcia2018chaos,rammensee2018many}. 
OTOCs have also given rise to other closely related measures~\cite{kukuljan2017weak,rozenbaum2017lyapunov,modi.2022}. Relating OTOC with quantum information theoretic notions like entanglement generation, uncertainty relations, discord, and quasi-probability distribution implies a closer tie between the tools of quantum information theory and quantum chaos~\cite{lewis2019unifying,alba2019quantum,styliaris2021information,seshadri2018tripartite,madhok2018quantum,madhok2015signatures}.

The validity of thermodynamic laws of motion from the statistical mechanical standpoint inherits the system's chaotic behaviour. Latora and Baranger proposed \cite{latora1999kolmogorov} a form of entropy similar to observational entropy (see next paragraph) that we study in this work, called ``physical entropy'', and extracted the Kolmogorov exponents for various simple classical chaotic maps. The foundations of quantum statistical mechanics rely on the concepts of quantum chaos, and the relation between OTOC and thermalization is well-studied in various models~\cite{gogolin2016equilibration,goold2016role}. The development of quantum information theory inherited all the notions of classical information theory by treating quantum mechanics as a generalization of probability theory into the non-commutative world, with various definitions of entropy generalized to quantum entropies. Still, what is to be considered thermodynamic entropy in quantum mechanics (QM) remains a controversial topic. Von Neumann's entropy was a valid form of entropy to study equilibrium thermodynamics as a generalization of Gibb's entropy. The interpretation of it as thermodynamic entropy and other foundational issues of statistical thermodynamics remains unanswered~\cite{von2010proof, She99, HS06, MMB05, DRE+11, AY13}. 

Safranek, Deutsch, and Aguirre recently studied the thermalization of closed quantum systems by defining an entropy, a quantum mechanical generalization of classical Gibbs and Boltzmann entropy \cite{deutchquantum2019, quantum2020deautch}. They called it {\it observational entropy} (OE) and proved that it is a monotonic function of coarse-graining. However, the concept of  OE is quite old and was introduced earlier in many seminal works in some different forms\cite{gibbs1902elementary,ehrenfest1990conceptual,von2010proof, Caves_notes}. Various extensions and applications of OE have been studied recently \cite{vsafranek2021brief,strasberg2021second,vsafranek2020classical, Buscemi_22,schindler2020quantum,strasberg2021second}. The main advantage of observational entropy is that it can be realized in currently available experimental setups \cite{carr2009cold,safronova2018search,atature2018material,vandersypen2019semiconductor,hensgens2017quantum, 
hartmann2008quantum,vaidya2018tunable,norcia2018cavity,davis2019photon,wang2015topological,noh2016quantum,
hartmann2016quantum,kjaergaard2020superconducting,ganzhorn2019gate,kandala2017hardware,hempel2018quantum,nam2020ground,britton2012engineered,bohnet2016quantum}. Even from the theoretical perspective, observational entropy can be a very useful diagnostic tool to characterize different phases of matter. In a recent work, two authors of this article investigated the localization-delocalization transition using OE. In contrast to the other diagnostic tools,  OE possesses an extra degree of freedom: the coarse-graining length. One can find an optimal coarse-graining length so that the finite size scaling shows much better data-collapse, which other diagnostic tools cannot provide~\cite{modak2022observational}.   

Our main aim is to investigate the regular and chaotic behaviour using OE. More precisely,
we want to investigate how  OE behaves for a system whose classical counterpart possesses regular, mixed, and chaotic phases. One of the best candidates for this study is the quantum kicked top (QKT) model \cite{haake1991quantum, haake1987classical, peres1997quantum,chaudhury2009quantum,neill2016ergodic,dogra2019quantum,sreeram2021out}. Depending on the strength of the kicking parameter, the classical analogue of this model shows a regular, mixed, or chaotic phase. QKTs are also experimentally realisable in cold atom \cite{chaudhury2009quantum}, and superconducting systems \cite{neill2016ergodic}. The eigenvalue-spacing statistics of this model show a transition from Poisson to Wigner distribution as expected, depending on the underlying classical dynamics \cite{haake1987classical}. Dynamical measures, such as sensitivity to perturbation, OTOCs, and entanglement dynamics, correspond with the classical phase space \cite{chaudhury2009quantum, wang2004entanglement, stamatiou2007quantum, ghose2008chaos,prosen2003theory, dogra2019quantum, sreeram2021out,schack1994hypersensitivity,sahu2022effect}. Surprisingly, kicked tops even in the deep quantum regime show signatures of chaos \cite{chaudhury2009quantum,dogra2019quantum,ghose2008chaos,sreeram2021out}.  Furthermore, such few-qubit QKTs are exactly solvable \cite{dogra2019quantum, sreeram2021out}, making them one of the few chaotic models with an analytical and experimental grasp.

\section{Formalism and model} 

\subsection{Observational entropy} 

Consider a quantum system $\rho$ defined on the Hilbert space $\mathcal{H}$ of dimension $d$. We can then partition $\mathcal{H}$ into orthogonal subspaces $\lbrace \mathcal{H}_i\rbrace$ such that  $ \mathcal{H} = \oplus_i \mathcal{H}_i$.  The projection operator onto a subspace $\mathcal{H}_i$ is denoted by $\Pi_i,$ and $\sum_i \Pi_i = \mathbb{I},$ since they form a complete set of projections. Such a set $\lbrace \Pi_i \rbrace$ is called a \textit{coarse-graining}, denoted by  $\chi$.  Each of the subspaces can be treated as a macrostate, and the probability $p_i$ that the system $\rho$ is found in a macrostate (subspace) $\mathcal{H}_i$ on measurement is given by $p_i = \mathrm{Tr} (\Pi_i \rho)$. Note that in general, both measurements and coarse-graining can be defined with general positive operator valued measures~\cite{safranek_gen}. The dimension of the subspace $\mathcal{H}_i$, given by  $ \mathrm{Tr} (\Pi_i)$ is called the \textit{volume} or \textit{coarse-graining length} ($V_i$)  of the subspace.
Then the observational entropy of the state $\rho$  associated with the coarse-graining $\chi$ is given as 
\begin{equation}
S_\chi (\rho) = -\sum_i p_i \log \frac{p_i }{V_i}. 
\label{eq:oe} 
\end{equation}

Consider two coarse-graining $\chi_1$ and $\chi_2$ with the projector sets $\Pi_{i_1}$ and $\Pi_{i_2}$. The coarse graining $\chi_1$ is {\it rougher} than the coarse-graining $\chi_2$, and is denoted  $\chi_1 \huk \chi_2,$ if for every $\Pi_{i_1} \in \chi_1,$ there exists $\lbrace \Pi_{i_2}\rbrace$ such that  $\Pi_{i_1} = \sum_{i_2 \in c_{i_1}} \Pi_{i_2}$, where $c_{i_1}$ is some index set. In this case, $\chi_2$ is called as {\it finer} coarse-graining than $\chi_1$. The coarse-graining $\chi_{\mathbb{I}}$ with idenity $\mathbb{I}$ is the roughest coarse graining,  as~ $\chi_{\mathbb{I}} \huk \chi_i$, for any coarse-graining $\chi_i$.  Coarse-graining with $\lbrace \Pi_{i_1}\rbrace$ containing only  rank-1 projectors ($V_i = 1,\: \forall i$) is the finest coarse-graining.

 It has been shown that  OE is a monotonic function of the coarse-graining. Given $\chi_1$ and $\chi_2$, if $\chi_1 \huk \chi_2$ then 
\begin{equation}
S_{\chi_1} (\rho) \geq S_{\chi_2} (\rho). 
\label{eq:oegrowth}
\end{equation}

For a given state of the system $\rho$, the von Neumann entropy $S_{vN}(\rho) = -\Tr [\rho \log \rho]$ bounds the OE $(S_{\chi}(\rho))$ for any coarse-graining $\chi$ 
\begin{equation}
    S_{vN}(\rho) \leq S_\chi (\rho) \leq \log \text{dim} \mathcal{H}. 
\end{equation}
However, note that the von Neumann entropy is invariant under the closed system (unitary) evolution. 
The empirical evidences and the laws of thermodynamics suggests that there are situations in which the thermodynamic entropy should increase in isolated system as well. 
This incites to propose a quantum analogue of thermodynamic entropy, and OE is seen as a natural alternative~\cite{quantum2020deautch,deutchquantum2019,strasberg2021second}. This is more evident if we rewrite the Eq.~(\ref{eq:oe}) as 
\begin{equation}
    S_\chi (\rho) = -\sum_i p_i \log p_i  + \sum_i p_i \log V_i.
\end{equation}
The first term is the Shannon entropy of measurement and the second term is the averaged Boltzmann entropy. The OE can have interpolations between both these entropies. The subspace $\mathcal{H}_i$ acts as a macrostate and if the state of the system $\rho$ is contained in one of the macrostates, then $\Pi_i \rho \Pi_i = \rho$. As a result, the  OE $S_\chi (\rho) = \log V_i$, which is precisely the quantum analogue of Boltzmann entropy. The Shannon entropy term indicates the mean uncertainty about the macrostate in which the system can be found. If the microstates in a given macrostate are indistinguishable, the Boltzmann term represents the mean uncertainty about the system after the measurement. If the system is chaotic, the associated uncertainty will have its representation in its dynamics, and the OE can capture it efficiently.   

{{To give an insight as to why OE diagnoses chaos, we consider OE from the perspective of retrodiction~\cite{buscemi2022observational,Busc_talk,Busc_retro}.}} The second law of thermodynamics and the thermodynamic fluctuation relations are derived from the difference between the prediction and retrodiction~\cite{buscemi2021fluctuation,aw2021fluctuation}.
 Consider a process represented by a random variable, which takes the value $x$ at time $t=0$, and $y$ at time $t=t'$.  We use the knowledge of the random variable  at $t=0$ to predict the value it takes at $t=t'$. Similarly, by using the knowledge of $y$ at $t=t',$ we can retrodict the value of the random variable at $t=0$.  The {\it predicting} probability is  $P_p (x,y) = p(x) p(y|x)$ and the {\it retrodicting} probability is $P_r (x,y) = q(y) q(x|y)$.   In the quantum mechanical scenario, let the quantum state in its eigen decomposition  $\rho = \sum_j \lambda_j \ketbra{\phi_j}$ be measured using the coarse-graining $\chi = \{\Pi_i \}$. Then the joint probability $P_p (i,j) = \lambda_j \expval{\Pi_i}{\phi_j}$. Using the measurement statistics of the coarse-graining $\chi = \{\Pi_i \} $, the retrodicted quantum state can be given as 
\begin{equation}
    \rho_{\text{rec}} := \sum_j \frac{p_j}{V_j} \Pi_j,
\end{equation}
the joint retrodictive probability is $P_r (i,j) = \left(p_j/V_j\right) \expval{\Pi_i}{\phi_j}$. The OE satisfies the following equality and an inequality in terms of the distance between the quantities that represent prediction and retrodiction, 
\begin{align}
    S_\chi (\rho) - S_{vN} (\rho) & = D_{KL} (P_p || P_r)  \\
    & \geq D (\rho || \rho_{\text{rec}}) . \label{rec}
\end{align}
Here $D_{KL} (\bullet||\bullet)$ is the classical relative entropy (Kullback-Leibler divergence)~\cite{kullback1951information,cover1999elements} and $D (\bullet||\bullet)$ is the Umegaki quantum relative entropy~\cite{umegaki1962conditional,wilde2013quantum}. 
For closed Hamiltonian dynamics (unitary dynamics), as pointed out earlier, the von Neumann entropy remains unchanged, and it vanishes for pure states. In some sense, chaos, residing in our understanding in its classical domain, represents the complex dynamics that relays on the difficulty to retrodict the initial state. {{Equation \ref{rec} conveys this intuition in the quantum realm. It says that a larger difference between the coarse-grained entropy(OE) and von Neumann entropy implies a bigger variance between the actual state of the system and the retrodicted system state from the coarse-grained measurements. Since von Neumann entropy is a constant for closed systems,  the $S_\chi (\rho)$ term (OE)  captures ignorance about the system because of coarse-grained measurements. We observe that ignorance about the system rapidly increases in quantum systems with classically chaotic dynamics. In other popular dynamical chaos measures such as Loschmidt echo and OTOC,  a perturbation in the dynamics depicts the ignorance of the system's interaction with the incompletely known environment. Both perturbation and coarse-graining represent a lack of complete knowledge about the system, and this manifests as a rapid  increase in system entropy in chaotic quantum systems. }} 

While the rate of entanglement entropy production is already used as a chaos indicator in the literature \cite{asplund2016entanglement,bianchi2018linear,lerose2020bridging},  having a classical analogue in physical entropy with a close relationship to the Kolmogorov-Sinai entropy-rate \cite{latora1999kolmogorov}, OE is a more natural candidate to diagnose chaos. 
Hence, in this work, we consider the QKT model and study the time evolution of $S_\chi (\rho)$ starting from initial localized spin coherent states. Having discussed the connection between OE and retrodiction,  intuitively, we expect 
$S_\chi (\rho)$ to grow faster and saturate to a larger value for chaotic dynamics than regular motions. Also,  Eq. (\ref{rec}) suggests that recovering spin-coherent states must be much more challenging in the chaotic regime. A recent article \cite{sahu2022effect} on spin coherent state tomography also corroborates this intuition. In the subsequent sections, we will discuss our findings in detail.


\subsection{ Quantum kicked top} 
We consider the QKT~\cite{haake1991quantum,haake1987classical} as the model for our study, and the Hamiltonian corresponding to QKT is given as 
\begin{equation}
H(t) = \frac{\hbar \alpha}{\tau} J_y + \frac{\hbar \kappa }{2j} J_z^2 \sum_{-\infty}^{\infty} \delta (t-n\tau).
\label{eq:qkt}
\end{equation}
The Hamiltonian consists of a rotation about the $Y$ axis, and periodic kicks about the $Z$ axis at time intervals, $\tau$.
   $\kappa$ is the kicking parameter and   $(J_x,J_y,J_z)$ are $x$, $y$, and $z$ components of the total angular momentum operators of spin $j$ system. The unitary operator corresponding to the QKT Hamiltonian Eqn.~(\ref{eq:qkt}) is 
\begin{equation}
U = \exp (-i \frac{k}{2j} J_z^2 ) \exp(-i \alpha  J_y). 
\label{eq:uniqkt}
\end{equation} 
 The dynamics of spin $J$ under the QKT unitary is given as $J_i^\prime = U^\dagger J_i U$. Given an initial state $\ket{\psi(0)}$, the  time evolved (discrete) state  $\ket{\psi(n)}$ under the QKT Hamiltonian is obtained by iterative application of the unitary operator $U$,  
\begin{equation}
\ket{\psi(n)} = U^n \ket{\psi(0)}.
\label{eq:uni}
\end{equation}

In the classical limit  $j\rightarrow \infty$ and  for $\alpha  = \pi/2$, by defining $X= \expval{\frac{J_x}{j}}, ~ Y= \expval{\frac{J_y}{j}} $, and $Z= \expval{\frac{J_z}{j}}$,  the maps takes a simple form as follows 
\begin{equation}
\begin{split}
X^\prime &= Z \cos (\kappa X) + Y \sin (\kappa X),\\
Y^\prime &= -Z \sin (\kappa X) + Y \cos (\kappa X), \\
Z^\prime &= -X.
\end{split}
\end{equation} 
 

\begin{figure*}
\includegraphics[scale=0.5]{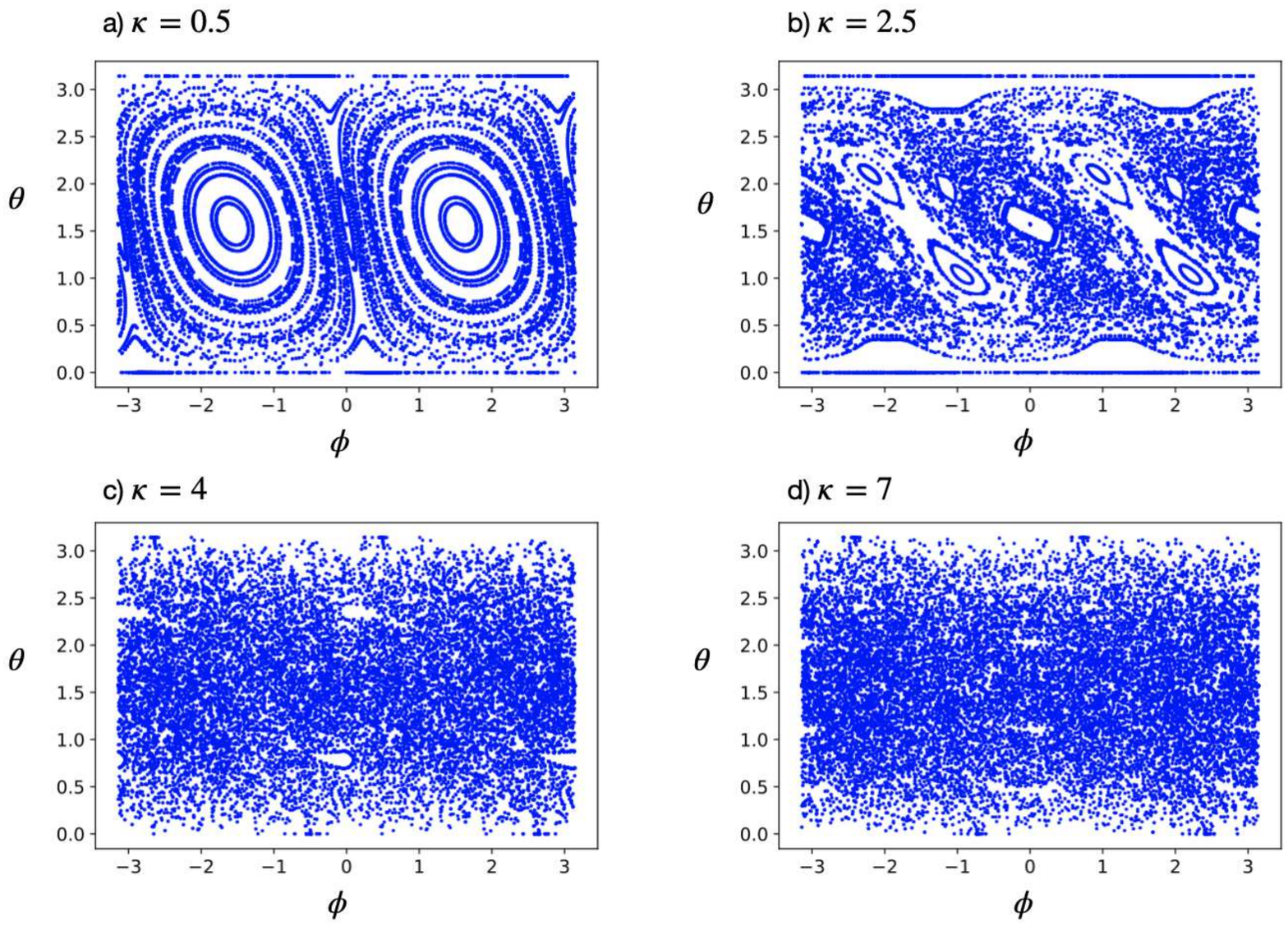}
\caption{Phase space distribution for various kicking strengths. At $\kappa=0.5$, the phase space is regular. At $\kappa=2.5$, the mixed phase space contains regular and chaotic regions. On increasing $\kappa$, the regular islands shrink and at $\kappa=7$, the phase space is completely chaotic. }
\label{fig:ckt}
\end{figure*} 
The classical map on the unit sphere for various values of the kicking strength $\kappa$ can be seen in Fig.~(\ref{fig:ckt}).  The visibly homogeneous blue region in Fig.(\ref{fig:ckt})c for instance, is called the chaotic region. The small structures floating in the chaotic sea are called regular regions. One can see that the regular regions gradually shrink and disappear as $\kappa$ increases in Fig. (\ref{fig:ckt}). Further details of the phase space for different $\kappa$ values can be found in \cite{haake1987classical}. The QKT~\cite{haake1991quantum,haake1987classical} is a simple quantum chaotic model studied from various perspectives~\cite{peres1997quantum,chaudhury2009quantum,neill2016ergodic,dogra2019quantum,sreeram2021out}.

\section{Kinematical study : Observational entropy with coarse-graining \label{sec:coa_obs}}

We study time evolution of spin coherent states with the QKT Hamiltonian (\ref{eq:qkt}) for various kicking parameters $\kappa$. The spin coherent states are defined as: 
\begin{equation}
 \ket{\psi(\theta,\phi)} = \frac{e^{\beta J_-}}{(1+\beta \beta^*)^j } \ket{j,j},
 \label{eq:coh} 
\end{equation}
where $\beta = e^{i\phi} \tan (\theta/2)$ and $J_- = J_x -i J_y$. The state $\ket{j,m}$ is the joint eigenstate of angular momentum operators $J^2$ and $J_z$ : 
\begin{equation}
 \begin{split}
  J^2 \ket{j,m} &= j(j+1) \ket{j,m} \\
  J_z \ket{j,m} &= m \ket{j,m}. 
  \label{eq:spinJ} 
 \end{split}
\end{equation}

The Hilbert space is of dimension $d = 2j+1 = 1024$, and the measurement operator is $J_z$. The eigenstates of $J_z$ are the computational basis vectors, denoted as $\lbrace \ket{q} \rbrace, ~ q \in \lbrace{0,1,... \:d-1 \rbrace}$. We construct the orthogonal subspace-projection operators from these computational basis vectors as follows. We define 
$\Pi_i=\sum_{q=i}^{i+k} \ket{q} \bra{q}$, where $i \in \lbrace 0,1,2... \:s-1\rbrace$ and $k <d$ is a constant.  Here $s$ denotes the total number of orthogonal partitions such that $\sum_{i=0}^{s-1} \Pi_i = \mathbb{I},$ and $\mathcal{H} = \oplus_{i=0}^{s-1} \mathcal{H}_i$, where $\mathcal{H}_i$ is the subspace onto which $\Pi_i$ projects the state. 
The dimensions of $\mathcal{H}_i$ are all the same, and the coarse-graining length $\mu = V_i$. 
Hence if $V_{i^\prime} > V_i$,
then $\chi_{i^\prime} \huk \chi_i$.  We consider the total  Hilbert space of dimension $d = 2j+1 = 1024,$ and study the growth of observational entropy with  coarse-graining length. We choose $\mu$ as integer powers of $2$.

\begin{figure}

 \includegraphics[scale=0.6]{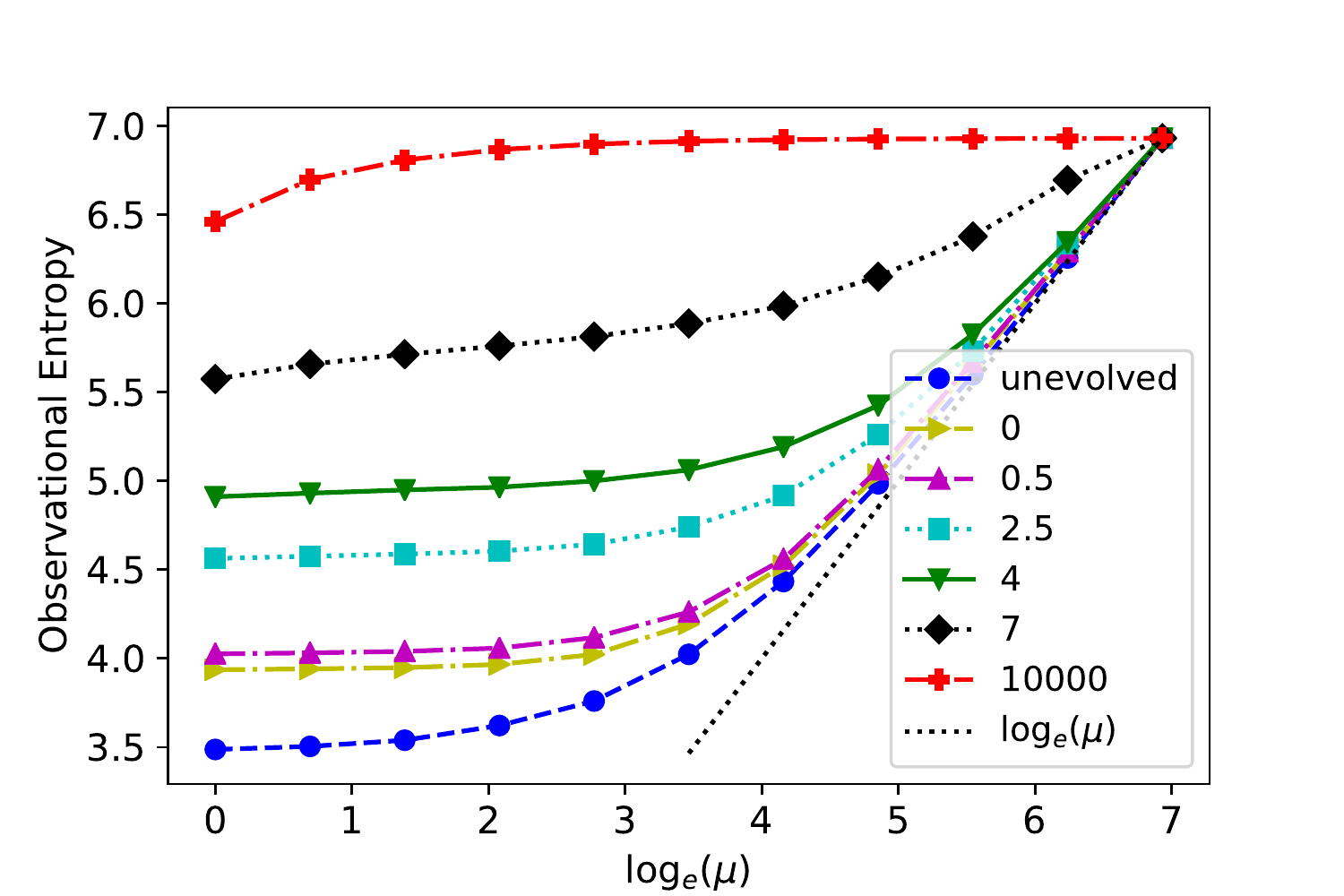}
 
 \caption{OE growth against natural log of the coarse-graining length ($\mu$) for different values of kicking parameter for $d=1024$. The Boltzmann term is given by $\log_e \mu$ for uniform coarse-graining. OE is dominated by the Boltzmann term for smaller chaoticity values at larger $\mu$. The growth of the OE with  $\mu,$ averaged over the initial states   we consider is denoted by ``unevolved". } 
 \label{fig:oecg}
\end{figure}

In Fig.~(\ref{fig:oecg}), we study the growth of OE with respect to the log of the coarse-graining length $\mu$. The initial state is the spin coherent state as defined in Eqn.~(\ref{eq:spinJ}), and the OE is calculated by averaging over $100$ states for uniformly chosen values of $\theta \in \{0,\pi\}$ and  $\phi \in \{0,2\pi\}$. The OE, as defined in Eqn.~(\ref{eq:oe}), consists of two terms. The first term is the Shannon entropy, and the second is the averaged Boltzmann entropy. Since we consider the volume of the subspaces $\mathcal{H}_i$ to be the same for every coarse-graining, the second term is a constant for a fixed $\mu$ and, i.e. $\log\mu$ which is shown in  Fig.~(\ref{fig:oecg}) using a dotted line. The growth of the OE with coarse-graining length $\mu$ averaged over initial states (without any time evolution) is also shown in the same figure.  After a critical value of coarse-graining length $\mu$, the OE grows as $\sim \log\mu$. It implies that the second term in the OE expression (averaged Boltzmann entropy term) starts dominating in this regime. On the other hand, for  higher kicking strengths, OE is already large even at small coarse-graining lengths, and OE does not grow as fast as $\mathrm{log}\:\mu$ anywhere. Note that the maximum attainable OE is $\log(\text{dim}(\mathcal{H}))$ which in our study is $\log (1024) \approx 6.93 $, and the approach towards the maximum value becomes faster with the increase in the kicking parameter strength.
Hence, the OE growth with coarse-graining length can clearly capture the distinctive behaviour of chaotic and regular motions.

\section{Dynamical study of Observational entropy and comparison with OTOC \label{sec:lya}}


For classical systems, chaos is characterized by how fast the trajectories can fill the entire phase space. For the QKT in its classical limit, one can see (from Fig.~(\ref{fig:ckt})) the spread of the trajectories in the phase space. This intuition is qualitatively expanded to calculate the Lyapunov exponents from the definition similar to the definition of OE as in Eqn.~(\ref{eq:oe}) for classical discrete maps in Ref. \cite{latora1999kolmogorov}. They could extract Lyapunov exponents from the slope of the growth of the physical entropy for various classical chaotic maps.



\begin{figure}
     \centering
     \begin{subfigure}[b]{0.5\textwidth}
         \centering
         \includegraphics[width=\textwidth]{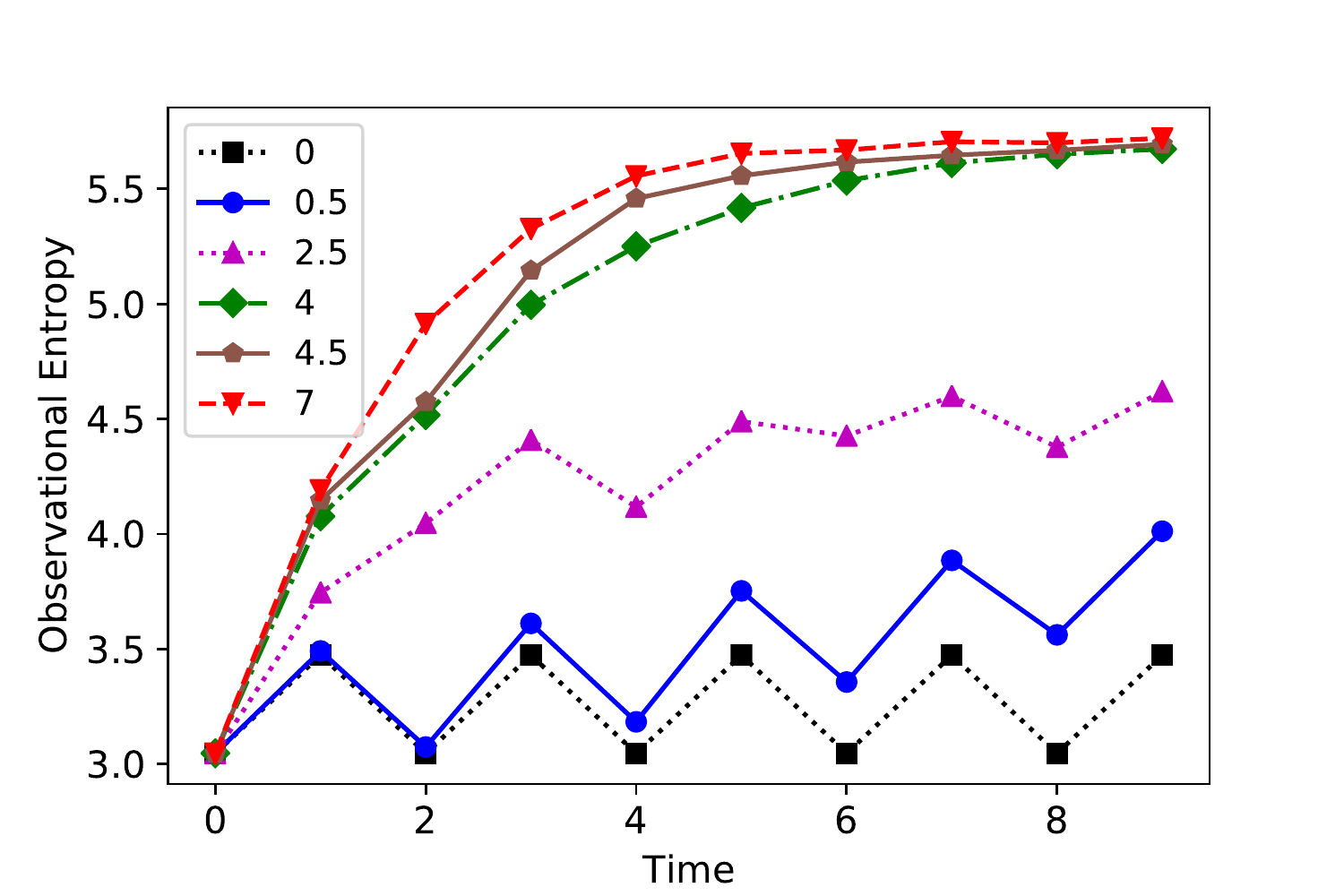}
         \caption{ }
        \label{fig:oete}
     \end{subfigure}
     \hfill
     \begin{subfigure}[b]{0.5\textwidth}
         \centering
         \includegraphics[width=\textwidth]{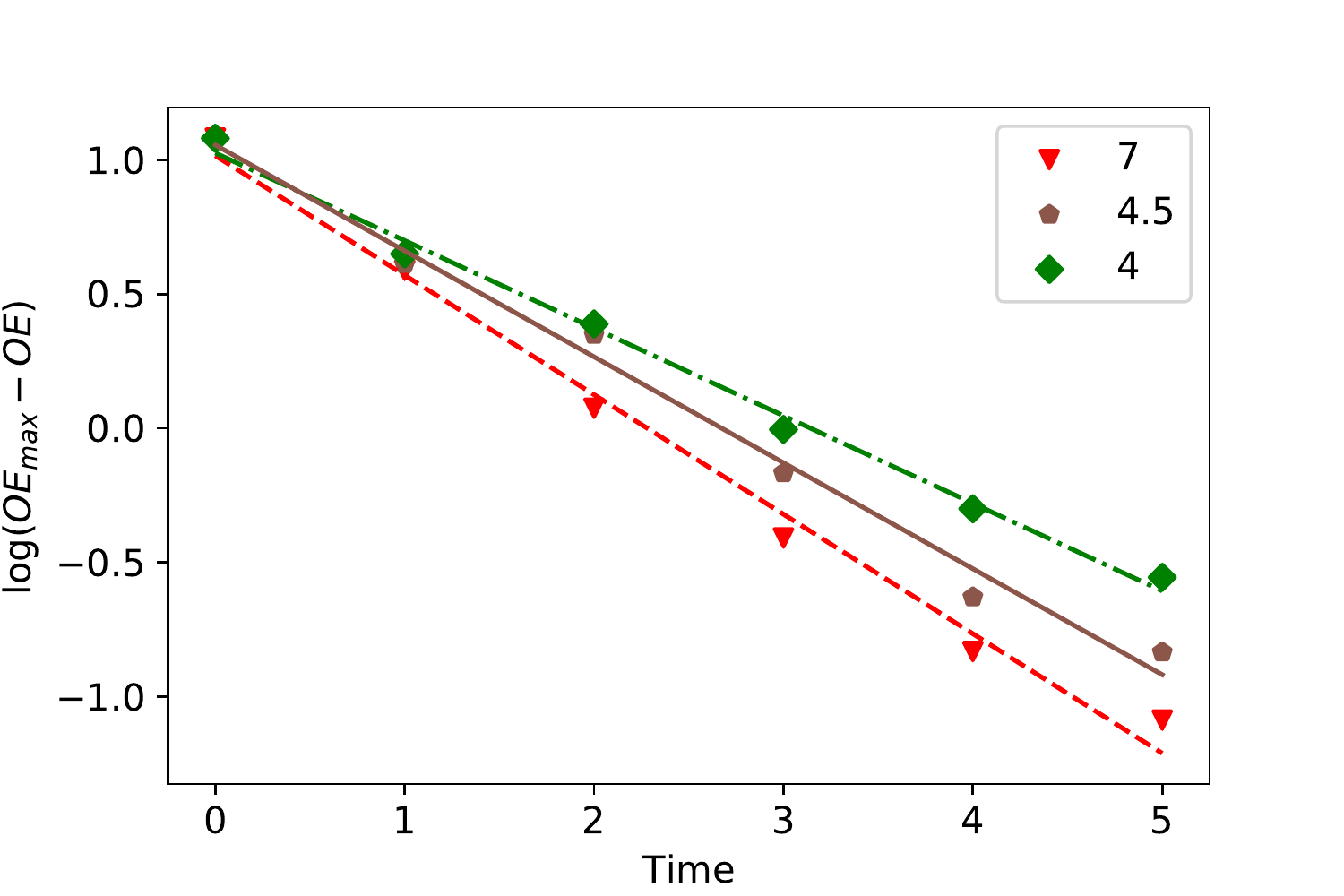}
         \caption{}
         \label{fig:oe_rate}
     \end{subfigure}
     
       \caption{(a). OE growth with respect to time for various kicking strengths $\kappa$. The dimension of the Hilbert space is $d = 400$. An average is taken over 100 coherent states. The rate of growth of OE  and its saturation value increases with $\kappa$. Recurrences can be observed in the regular regime. (b). Rate of increase of OE for $\kappa=4$, $\kappa=4.5$ and $\kappa=7$ in (a) shown in a semi-log plot.  $OE_{max}= \log 400$ is the maximum value attainable. Linearity of the curves observed till time step 5 implies an initial exponential increase towards saturation. The slope for $\kappa=7,\: 4.5$ and $4$ are $-0.445, -0.395$ and  $-0.326$ respectively.  }
        \label{fig:mix}
\end{figure}

The growth of OE for an initial state $\ket{\psi(\theta,\phi)}$  under the discrete time evolution $U$   for various kicking strengths $\kappa$ is studied in Fig.~(\ref{fig:oete}). Once again, we average over $100$ random initial coherent states to obtain the time evolution results. We choose the Hilbert space dimension $d = 2j+1 = 400$. Unlike the results shown in the previous section, here we choose a non-uniform coarse-graining (the coarse-graining length $\mu=2$ for half of the Hilbert space, and for the other half of the Hilbert space, we choose $\mu=4$). The reasons behind this particular choice of coarse-graining are twofold: 
1) OE shows a short time growth followed by saturation  in the chaotic regime. If one chooses $\mu$ to be large enough, the immediate consequence will be that the value of OE for the initial state will also be reasonably large. Hence, especially for higher kicking strengths, where the growth is  exponential, as shown in Fig. (\ref{fig:oe_rate}), the dynamical range of the short-time growth of OE will be extremely small for the Hilbert space dimension $d=400$. That will make our analysis inefficient. 
2) The second term (Boltzmann entropy) of the OE expression makes OE different from the usual Shannon entropy. Hence, to make the second term non-zero and keep the coarse-graining length small enough, the ideal choice is $\mu=2$. However, 
the uniform coarse-graining implies that the second term of the OE expression is just a constant, i.e. $\log 2$ (if $\mu$ is set to be 2) irrespective of the dynamics. Then OE does not contain more information about the evolving state than Shannon entropy. Hence, we choose a non-uniform coarse-graining and restrict ourselves to the coarse-graining length $\mu=2$ for half of the Hilbert space and  $\mu=4$ for the other half. However, we have also verified our results for other coarse-graining lengths, and our finding is qualitatively robust as long as the $\mu<< d$.

For small values of kicking parameters $\kappa$, the growth is slow and 
the long time saturation value is also much smaller than the maximum value, i.e. $\log(400)\simeq 5.99$. In contrast, for the higher kicking strengths, OE grows faster and reaches very close to the maximum value in the long time limit. Also, for the regular system (small values of $\kappa$), one can see the revivals in OE dynamics. On the other hand, in the chaotic case, the revival is not seen. These revivals are a signature of the existence of the regular (periodic) orbits in the classical phase space for such systems. Interestingly, 
similar revivals (or lack of it)  has also been observed for an integrable (non-integrable) quantum spin chain in the entanglement dynamics~\cite{modak2020entanglement}.

The initial rate of growth in Fig.~(\ref{fig:oete}) clearly distinguishes regular dynamics from a chaotic one. The growth rate is highest at initial time steps and later flattens to a saturation value in the chaotic regime.  We plot this initial period of growth for the chaotic regime separately in Fig. (\ref{fig:oe_rate}. Linear behaviour in the semi-log plot reflects an exponential growth of OE. The exponent increases with chaoticity.

\begin{figure}
 \includegraphics[scale=0.6]{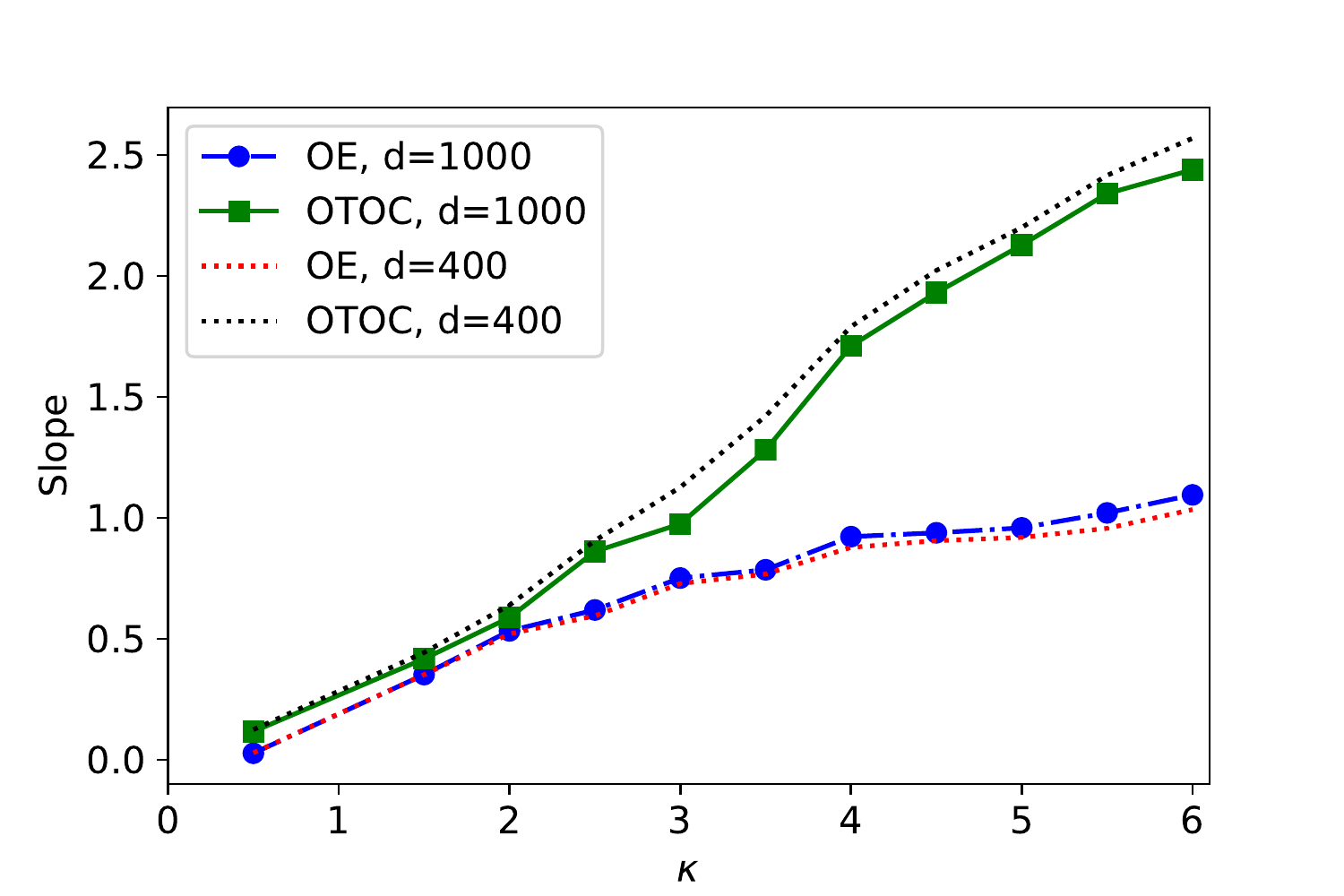}
 \caption{Initial rate of growth of OE and that of the $\log$ of OTOC, calculated at third time step is plotted against chaoticity, for $d=400$ and $d=1000$.    A linear fit yields the following slopes. OE, $d=1000:0.09182$, OE, $d=400:  0.08560$, which are very similar numbers. Also for OTOC, we get  OTOC, $d=1000: 0.24361 $ and OTOC, $d=400:  0.25176.$. Therefore, OE and OTOC rates of growth are very similar, despite the difference in dimensions of the Hilbert space.}
 \label{fig:oe_otocj}
\end{figure}

How does the growth rate change with the dimension of the Hilbert space? We see in Fig. (\ref{fig:oe_otocj}) that the growth rate of both  OE and OTOC are very similar at $d=400$ and $d=1000$. 
On increasing the chaoticity further, both the OTOC  and OE growth saturate (not shown in the figure), pertaining to the finite size of the Hilbert space.

 Next, we focus on the short time dynamics of the OE in detail.
 We define the OE growth rate $\lambda_{OE}$ as the slope of the short-time OE growth and compare it with that of OTOC, a popular diagnostic of chaos of late. OTOC quantifies the scrambling and the spread of initially localized quantum information. For an Hermitian operator $A(t)$, we define the OTOC as 
\begin{equation}
    C(t) = - \frac{1}{2} \Tr(\rho[A(t),A(0)]^2). 
    \label{eq:otoc}
\end{equation}
where $A(t) = U^\dagger(t) A U(t) $, time evolved operator under Heisenberg picture,  and $\rho$ is the initial state.   The rate of growth of OTOC, $\lambda_q$, associated with OTOC as 
\begin{equation}
    C(t) \approx e^{2\lambda_q t }.
    \label{eq:otocrate}
\end{equation}
 OTOC acts an indicator of the extent of chaos. For our system, we consider $A = J_z$, a spin operator and evolve under the unitary operator $U$ in Eqn.~(\ref{eq:uniqkt}). An average is taken over the $100$ states chosen randomly from the coherent states as described earlier. 
  First, we compare our $\lambda_{OE}$ (chaos indicator obtained from the short-time growth rate of OE) with  $\lambda_{q}$ 
 (chaos indicator obtained from OTOC) in Fig.~(\ref{fig:oe_otocj}) for $d=400$ and ${d=1000}$. Remarkably, we find that the behaviour $\lambda_{OE}$ and $\lambda_{q}$ is very similar for different values of $\kappa$. It is a validation that, indeed, 
 the short-time growth rate of OE can act as a diagnostic tool to quantify the chaotic behaviour in a quantum system.

\begin{figure}
 \includegraphics[scale=0.6]{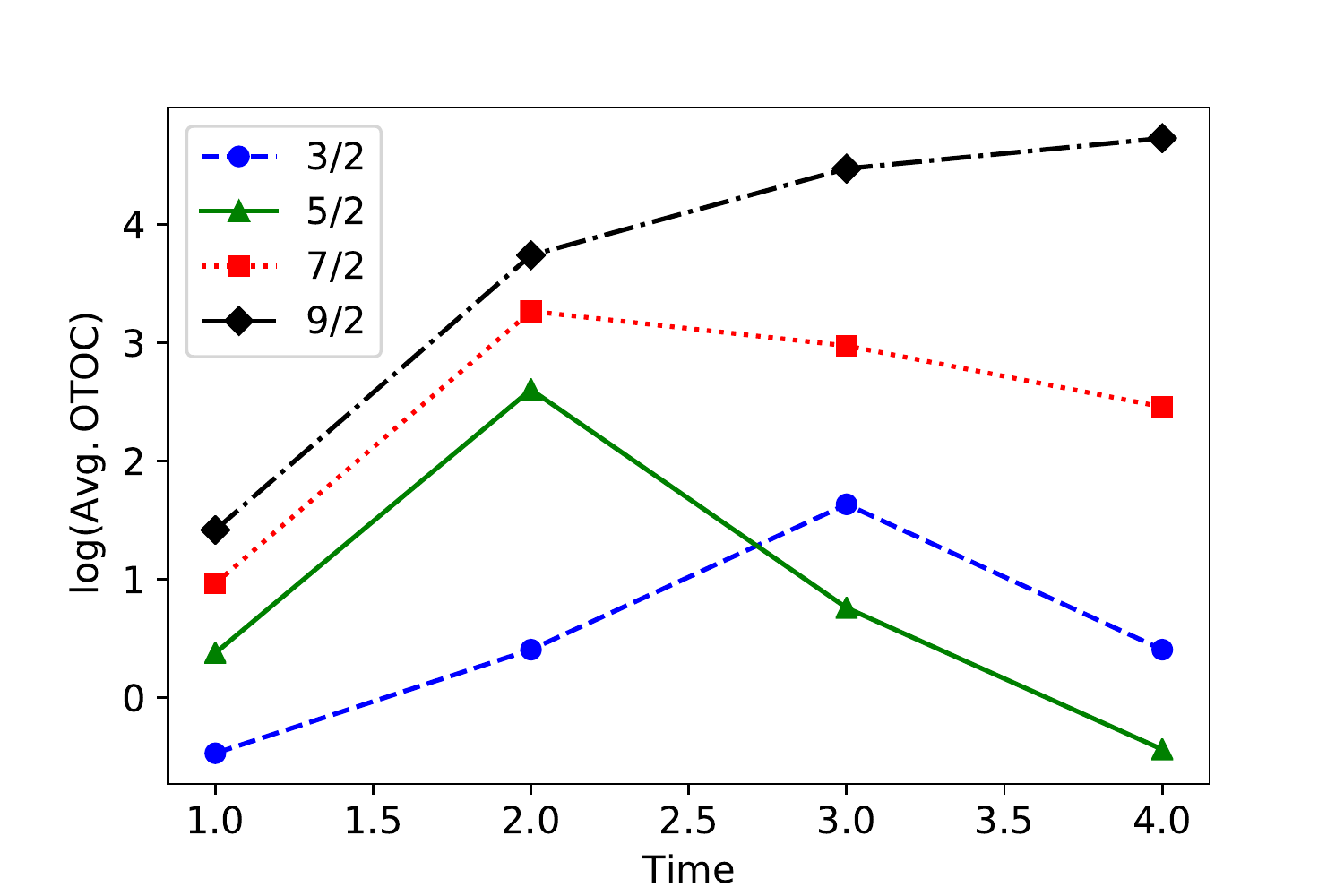}
 \caption{Growth of OTOC with time for small $j$ values in the fully chaotic regime, $\kappa=3\pi/2$. Saturation of the initial rate of growth can be observed at $j=5/2.$ The average is taken over 100 coherent states as explained in the main text. }
 \label{fig:otocsmallj}
\end{figure}

  We ask the following question: how well does OE capture the signatures of chaos in the deep quantum regime (small values of $j$)? OTOC is a sensitive diagnostic tool that has detected vestiges of chaos in kicked top models consisting of three and four qubits \cite{sreeram2021out}. The small $j$ behaviour obtained in \cite{sreeram2021out} is reproduced in Fig. (\ref{fig:otocsmallj}).   Exponential rise in OTOC is expected till Ehrenfest time. The Ehrenfest time  $t_E  \sim \frac{\log (1/ h_{eff}) }{\log (\kappa/2)}  \sim \frac{\log (2J+1) }{\log (\kappa/2)}  $ \cite{rozenbaum2017lyapunov}.
 For our choices of $\kappa$ in the chaotic regime and small $J$ values, the above formula gives $\frac{\log (2J+1) }{\log (\kappa/2)} \sim 1$.  Since the Ehrenfest time for these models is extremely short, and the observation is confined to the first two time steps. The initial rate of growth of OTOC saturates at $j=5/2$, and higher quantum numbers exhibit the same slope between the first two time steps. This indicates that to witness the chaotic growth rate, one needs to go only as high as the quantum number $j=5/2$. Interestingly, Loschmidt echo \cite{peres1984stability,jalabert2001environment}, another quantifier of chaos, did not show such sensitivity, and it took considerably larger angular momentum to show an exponential decay, indicating the underlying chaos \cite{sreeram2021out}.

\begin{figure}
 \includegraphics[scale=0.6]{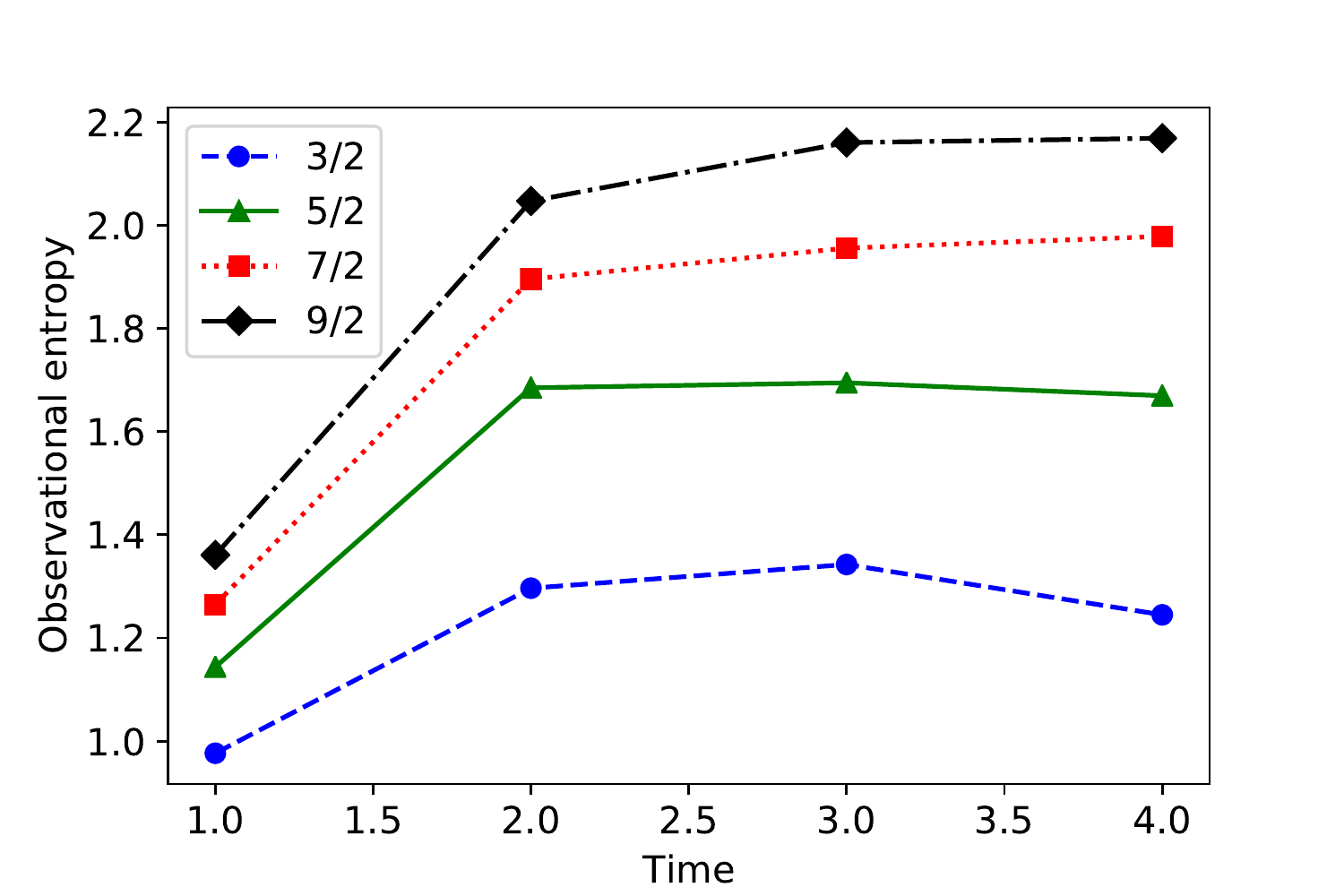}
 \caption{Growth of OE with time for small $j$ values in the fully chaotic regime of $\kappa=3\pi/2$.  Growth takes place  up to Ehrenfest time, and saturation occurs afterwards, even for the small $j$ values considered. This is unlike the OTOC behaviour.  The average is taken over 100 coherent states, as explained in the main text. }
 \label{fig:oesmallj}
\end{figure}

  Against this backdrop, we look at the OE growth for small $j$. We see in Fig.~(\ref{fig:oesmallj}) that OE shows an initial rise, and then it saturates for $j=7/2$ and $9/2$ at $\kappa=3\pi/2$, similar to its behaviour at larger quantum numbers. The growth of OE takes place in the pre-Ehrenfest regime, within the first two time steps. Interestingly, OTOCs show revivals for small $j$ values, as seen in Fig.~(\ref{fig:otocsmallj}). Post-Ehrenfest saturation in OTOCs is observed only at larger angular momenta. Hence, we conclude that the short-time growth of OE is a much more robust diagnostic tool to detect the degree of the chaoticity in a system even if $j$ is small. 


 \label{fig:OE_saddle}

\begin{figure}[h!]
     \centering
     \begin{subfigure}[b]{0.5\textwidth}
         \centering
         \includegraphics[width=\textwidth]{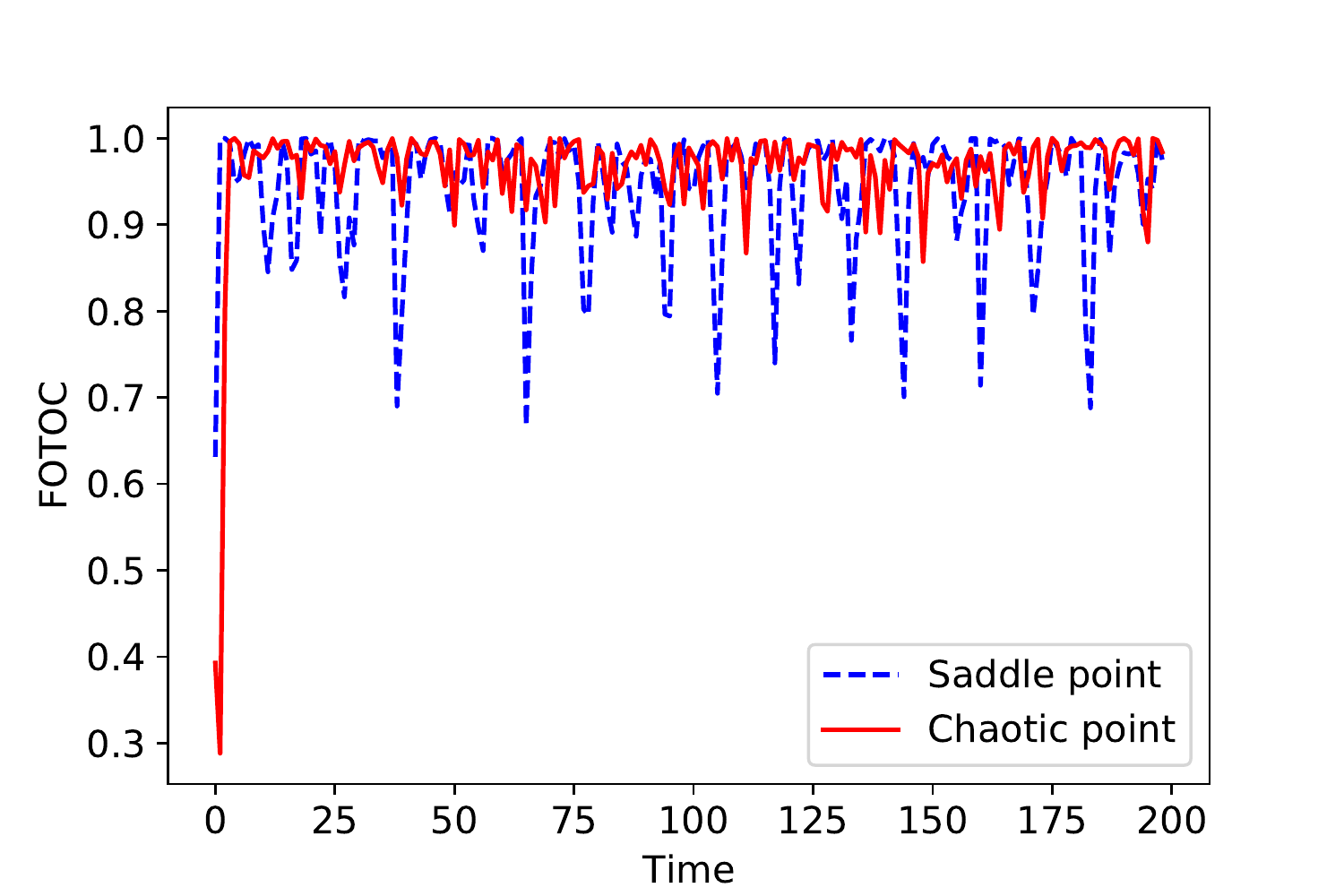}
         \caption{ }
        \label{fig:FOTOC_longtime}
     \end{subfigure}
     \hfill
     \begin{subfigure}[b]{0.5\textwidth}
         \centering
         \includegraphics[width=\textwidth]{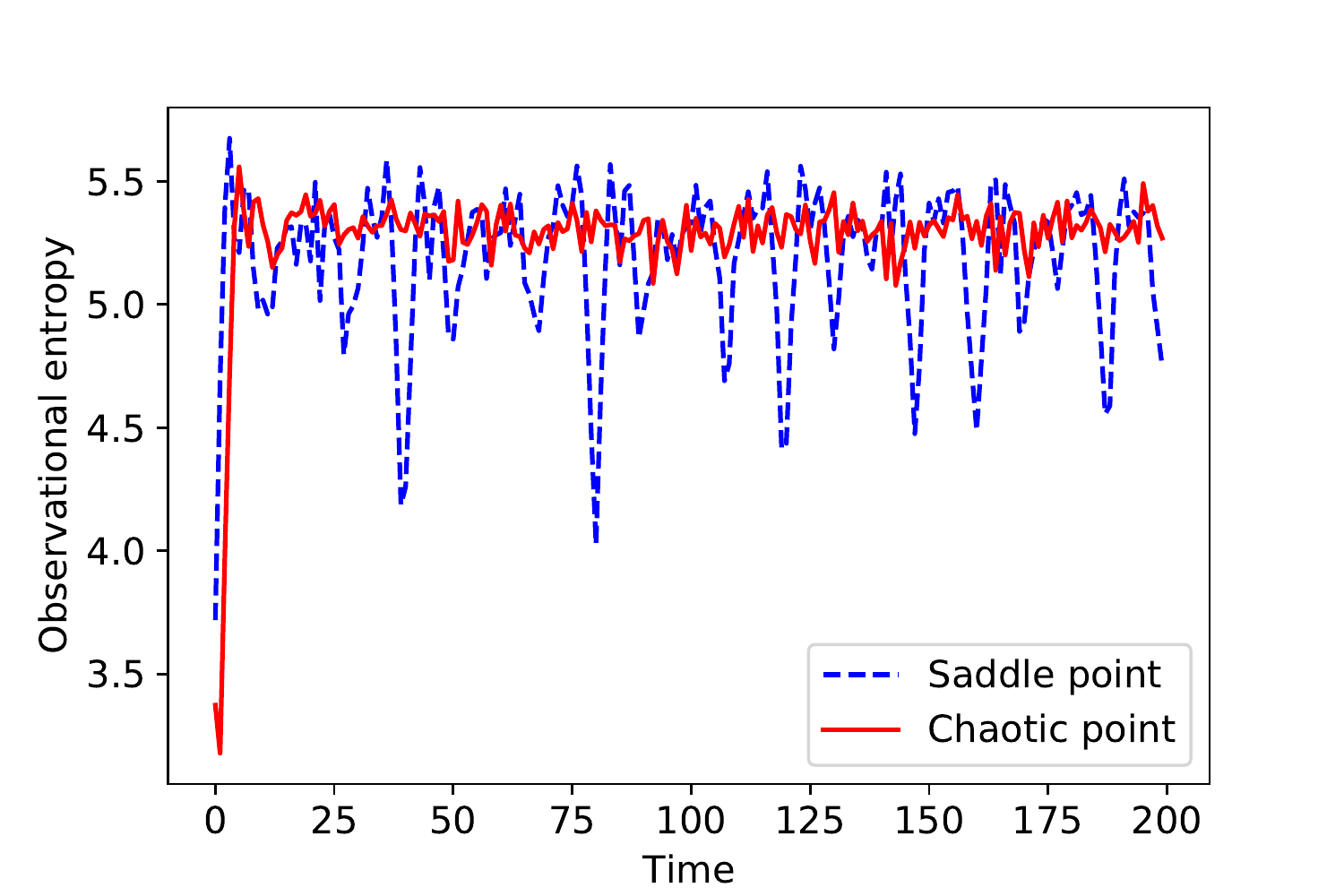}
         \caption{}
        \label{fig:oe_longtime}
     \end{subfigure}
     \caption{Saddle point behaviour vs chaotic initial states of FOTOC (a) and OE (b) at $\kappa=2.5$. The saddle point is situated in the classical phase space at $(\pi/2,\pi/2)$, and the chaotic point is chosen at $(\pi/4,\pi/4).$  Saddle point shows larger oscillations in both subfigures. Long-time dynamics of OE and FOTOC can clearly distinguish a chaotic point from a saddle in the mixed phase space.}
     \end{figure}
While the short-time behaviour of OE and OTOC can be used to identify chaos in a quantum system, it can sometimes be deceptive. For example, OTOC can grow exponentially even when the classical counterpart of the system is not chaotic. The presence of local instabilities, like a saddle point, can mimic chaos-like behaviour in OTOCs \cite{xu2020does}. Therefore, interpreting the exponential growth of OTOC with chaos is questionable.

The key to solving this problem is to look at the long-term dynamics. Kidd \textit{et al.} recently demonstrated that studying the long-time behaviour of OTOCs can distinguish true chaos from saddle-dominated scrambling \cite{kidd2021saddle}. Their numerical study involved the Bose-Hubbard dimer and a longer spin-chain model called the Dicke model. Using fidelity OTOCs (FOTOC), they showed that the expected saturation and convergence of the OTOC value is only seen in the chaotic regime. The post-Ehrenfest time behaviour of the saddle-dominated regime showed large oscillations, visibly distinguishing from true chaos.
Can OE reliably discern local instabilities? For OE to qualify as a genuine chaos indicator, it must pass this test. 

At $\kappa=2.5$, the kicked top classical phase space is mixed, with regular and chaotic regions coexisting and admitting a saddle point at $(\pi/2,\pi/2)$. We study the FOTOC behaviour for the saddle point and a chaotic point,  and compare it with OE dynamics.
FOTOC is defined as follows\cite{garttner2017measuring,garttner2018relating}.
\begin{equation}
    \mathrm{FOTOC}= 1- \mathrm{Re}\langle \hat{W}_\delta^\dagger (t) \hat{V}^\dagger(0) \hat{W}_\delta(t) \hat{V}(0) \rangle
    \label{FOTOC}
\end{equation}
We choose $\hat V$  as a coherent state projection operator, $\ketbra{\psi(\theta,\phi)}{\psi(\theta,\phi)}$, with the expectation in Eqn.~(\ref{FOTOC}) taken with respect to the same coherent state.  $\hat W_\delta (0)$ is a perturbation, modeled as a small-$\delta$ rotation about $X$ axis. The time evolution   $\hat W_\delta (t)=\hat U^\dagger (t) \hat W_\delta (0) \hat U(t)$ is governed by the kicked-top floquet. In this case, Eqn.~(\ref{FOTOC}) reduces to
\begin{equation}
    \mathrm{FOTOC}= 1- |\bra{\psi(\theta,\phi)}\hat W_\delta (t) \ket{\psi(\theta,\phi)}|^2.
\end{equation}
The long-time behavior of FOTOC is plotted in Fig. (\ref{fig:FOTOC_longtime}).
 It shows that a state $\ket{\psi(\theta,\phi)}$ located at the saddle point behaves quite distinctly from a state $(\pi/4,\pi/4),$ located in the chaotic region. The latter shows an exponential rise and smaller but persistent fluctuations around the saturation value. In contrast, the saddle-point FOTOC leads to quantitatively larger oscillations post-Ehrenfest time, indicating near revivals of the initial state. The persistent fluctuations in the chaotic case suggest that the system does not entirely thermalize in the time frame considered \cite{kidd2021saddle}.
Finally, we also study the long-time dynamics of observational entropy, starting from the saddle point $(\pi/2,\pi/2)$, and the same chaotic point $(\pi/4,\pi/4)$. Figure (\ref{fig:oe_longtime}) shows that the behaviour is qualitatively very similar to that of FOTOC, indicating that OE is as good as FOTOC in distinguishing chaos from saddle-dominated scrambling.

\section{Conclusions}
 In this work, we demonstrated how the OE behaves in the regular, mixed, and chaotic regime using a prototype model called quantum kicked top. First, we studied the variation of OE with coarse-graining length. After a critical coarse-graining length, we found that the Boltzmann term in the OE expression starts dominating in the regular regime. OE growth is logarithmic in coarse-graining length. On the other hand, in the chaotic regime,  OE growth is much faster. Next, we focused on the dynamics and demonstrated that the short-time growth rate of OE can be used as a measure of the chaoticity in the system and compared our results with OTOC.

Moreover, we showed that in the deep quantum regime, the results obtained from OE are much more robust than the OTOC results, making OE a superior candidate over OTOC to identify chaos in quantum systems. Finally, we also investigated the long-time behaviour of OE to distinguish between saddle point-driven scrambling from chaotic scrambling. We found that the saddle point OE shows large, persistent fluctuations compared to the chaotic regime, which has also been observed in the FOTOC study. Further, it will be interesting to investigate the OE of finite quantum spin chains~\cite{modak.2014} that shows a crossover between integrability and non-integrability but has no classical counterpart, unlike the QKT model.

 The emergence of the irreversible macroscopic world from the reversible laws of the microscopic world is an important issue both at its fundamental level and from the technological forefront~\cite{gogolin2016equilibration, d2016quantum, GHR+16,yukalov2011equilibration}. The connection between the many-body quantum chaos and the foundations of statistical mechanics in terms of equilibration and thermalization is well known~\cite{d2016quantum}. The OE can be expressed as a distance  between prediction and retrodiction~\cite{buscemi2022observational}, which provides a quantitative base for the fluctuations relations and the laws of thermodynamics~\cite{buscemi2021fluctuation,GHR+16,seifert2012stochastic}. More studies in OE would reveal the deeper correlation between the laws of statistical thermodynamics and quantum chaos. As OE involves only local measurements, it can be easily experimentally implemented in current quantum computing hardware~\cite{carr2009cold,safronova2018search,atature2018material,vandersypen2019semiconductor,hensgens2017quantum,
hartmann2008quantum,vaidya2018tunable,norcia2018cavity,davis2019photon,wang2015topological,noh2016quantum,
hartmann2016quantum,kjaergaard2020superconducting,ganzhorn2019gate,kandala2017hardware,hempel2018quantum,nam2020ground,britton2012engineered,bohnet2016quantum}, and can be potentially used to study the destabilizing effect of chaos in quantum device control.
 
\section{Acknowledgements}
RM acknowledges the DST-Inspire fellowship by the Department of Science and Technology, Government of India, SERB start-up grant (SRG/2021/002152). SA acknowledges the start-up research grant from
SERB, Department of Science and Technology, Govt. of India (SRG/2022/000467). Authors thank M.S Santhanam,  Bharathi Kannan, Abinash Sahu, Naga Dileep Varikuti and  Dominik {\v{S}}afr{\'a}nek for useful discussions.  
\bibliography{thermo,chaos} 

\begin{thebibliography}{102}%
\makeatletter
\providecommand \@ifxundefined [1]{%
 \@ifx{#1\undefined}
}%
\providecommand \@ifnum [1]{%
 \ifnum #1\expandafter \@firstoftwo
 \else \expandafter \@secondoftwo
 \fi
}%
\providecommand \@ifx [1]{%
 \ifx #1\expandafter \@firstoftwo
 \else \expandafter \@secondoftwo
 \fi
}%
\providecommand \natexlab [1]{#1}%
\providecommand \enquote  [1]{``#1''}%
\providecommand \bibnamefont  [1]{#1}%
\providecommand \bibfnamefont [1]{#1}%
\providecommand \citenamefont [1]{#1}%
\providecommand \href@noop [0]{\@secondoftwo}%
\providecommand \href [0]{\begingroup \@sanitize@url \@href}%
\providecommand \@href[1]{\@@startlink{#1}\@@href}%
\providecommand \@@href[1]{\endgroup#1\@@endlink}%
\providecommand \@sanitize@url [0]{\catcode `\\12\catcode `\$12\catcode
  `\&12\catcode `\#12\catcode `\^12\catcode `\_12\catcode `\%12\relax}%
\providecommand \@@startlink[1]{}%
\providecommand \@@endlink[0]{}%
\providecommand \url  [0]{\begingroup\@sanitize@url \@url }%
\providecommand \@url [1]{\endgroup\@href {#1}{\urlprefix }}%
\providecommand \urlprefix  [0]{URL }%
\providecommand \Eprint [0]{\href }%
\providecommand \doibase [0]{http://dx.doi.org/}%
\providecommand \selectlanguage [0]{\@gobble}%
\providecommand \bibinfo  [0]{\@secondoftwo}%
\providecommand \bibfield  [0]{\@secondoftwo}%
\providecommand \translation [1]{[#1]}%
\providecommand \BibitemOpen [0]{}%
\providecommand \bibitemStop [0]{}%
\providecommand \bibitemNoStop [0]{.\EOS\space}%
\providecommand \EOS [0]{\spacefactor3000\relax}%
\providecommand \BibitemShut  [1]{\csname bibitem#1\endcsname}%
\let\auto@bib@innerbib\@empty
\bibitem [{\citenamefont {Casati}\ and\ \citenamefont
  {Ford}(1979)}]{casati1979stochastic}%
  \BibitemOpen
  \bibfield  {author} {\bibinfo {author} {\bibfnamefont {Giulio}\ \bibnamefont
  {Casati}}\ and\ \bibinfo {author} {\bibfnamefont {Joseph}\ \bibnamefont
  {Ford}},\ }\bibfield  {title} {\enquote {\bibinfo {title} {Stochastic
  behavior in classical and quantum hamiltonian systems.[como, 1977]},}\
  }\href@noop {} {\  (\bibinfo {year} {1979})}\BibitemShut {NoStop}%
\bibitem [{\citenamefont {Shepelyansky}(1983)}]{shepelyansky1983some}%
  \BibitemOpen
  \bibfield  {author} {\bibinfo {author} {\bibfnamefont {Dima~L}\ \bibnamefont
  {Shepelyansky}},\ }\bibfield  {title} {\enquote {\bibinfo {title} {Some
  statistical properties of simple classically stochastic quantum systems},}\
  }\href@noop {} {\bibfield  {journal} {\bibinfo  {journal} {Physica D:
  Nonlinear Phenomena}\ }\textbf {\bibinfo {volume} {8}},\ \bibinfo {pages}
  {208--222} (\bibinfo {year} {1983})}\BibitemShut {NoStop}%
\bibitem [{\citenamefont {Berry}\ \emph {et~al.}(1979)\citenamefont {Berry},
  \citenamefont {Balazs}, \citenamefont {Tabor},\ and\ \citenamefont
  {Voros}}]{berry1979quantum}%
  \BibitemOpen
  \bibfield  {author} {\bibinfo {author} {\bibfnamefont {Michael~V}\
  \bibnamefont {Berry}}, \bibinfo {author} {\bibfnamefont {Nandor~L}\
  \bibnamefont {Balazs}}, \bibinfo {author} {\bibfnamefont {Michael}\
  \bibnamefont {Tabor}}, \ and\ \bibinfo {author} {\bibfnamefont {Andr{\'e}}\
  \bibnamefont {Voros}},\ }\bibfield  {title} {\enquote {\bibinfo {title}
  {Quantum maps},}\ }\href@noop {} {\bibfield  {journal} {\bibinfo  {journal}
  {Annals of Physics}\ }\textbf {\bibinfo {volume} {122}},\ \bibinfo {pages}
  {26--63} (\bibinfo {year} {1979})}\BibitemShut {NoStop}%
\bibitem [{\citenamefont {Dittrich}\ and\ \citenamefont
  {Graham}(1990)}]{dittrich1990long}%
  \BibitemOpen
  \bibfield  {author} {\bibinfo {author} {\bibfnamefont {T}~\bibnamefont
  {Dittrich}}\ and\ \bibinfo {author} {\bibfnamefont {R}~\bibnamefont
  {Graham}},\ }\bibfield  {title} {\enquote {\bibinfo {title} {Long time
  behavior in the quantized standard map with dissipation},}\ }\href@noop {}
  {\bibfield  {journal} {\bibinfo  {journal} {Annals of Physics}\ }\textbf
  {\bibinfo {volume} {200}},\ \bibinfo {pages} {363--421} (\bibinfo {year}
  {1990})}\BibitemShut {NoStop}%
\bibitem [{\citenamefont {Haake}(1991)}]{haake1991quantum}%
  \BibitemOpen
  \bibfield  {author} {\bibinfo {author} {\bibfnamefont {Fritz}\ \bibnamefont
  {Haake}},\ }\bibfield  {title} {\enquote {\bibinfo {title} {Quantum
  signatures of chaos},}\ }in\ \href@noop {} {\emph {\bibinfo {booktitle}
  {Quantum Coherence in Mesoscopic Systems}}}\ (\bibinfo  {publisher}
  {Springer},\ \bibinfo {year} {1991})\ pp.\ \bibinfo {pages}
  {583--595}\BibitemShut {NoStop}%
\bibitem [{\citenamefont {Hayden}\ and\ \citenamefont
  {Preskill}(2007)}]{hayden2007black}%
  \BibitemOpen
  \bibfield  {author} {\bibinfo {author} {\bibfnamefont {Patrick}\ \bibnamefont
  {Hayden}}\ and\ \bibinfo {author} {\bibfnamefont {John}\ \bibnamefont
  {Preskill}},\ }\bibfield  {title} {\enquote {\bibinfo {title} {Black holes as
  mirrors: quantum information in random subsystems},}\ }\href@noop {}
  {\bibfield  {journal} {\bibinfo  {journal} {Journal of high energy physics}\
  }\textbf {\bibinfo {volume} {2007}},\ \bibinfo {pages} {120} (\bibinfo {year}
  {2007})}\BibitemShut {NoStop}%
\bibitem [{\citenamefont {Shenker}\ and\ \citenamefont
  {Stanford}(2014)}]{shenker2014black}%
  \BibitemOpen
  \bibfield  {author} {\bibinfo {author} {\bibfnamefont {Stephen~H}\
  \bibnamefont {Shenker}}\ and\ \bibinfo {author} {\bibfnamefont {Douglas}\
  \bibnamefont {Stanford}},\ }\bibfield  {title} {\enquote {\bibinfo {title}
  {Black holes and the butterfly effect},}\ }\href@noop {} {\bibfield
  {journal} {\bibinfo  {journal} {Journal of High Energy Physics}\ }\textbf
  {\bibinfo {volume} {2014}},\ \bibinfo {pages} {1--25} (\bibinfo {year}
  {2014})}\BibitemShut {NoStop}%
\bibitem [{\citenamefont {Maldacena}\ \emph {et~al.}(2016)\citenamefont
  {Maldacena}, \citenamefont {Shenker},\ and\ \citenamefont
  {Stanford}}]{maldacena2016bound}%
  \BibitemOpen
  \bibfield  {author} {\bibinfo {author} {\bibfnamefont {Juan}\ \bibnamefont
  {Maldacena}}, \bibinfo {author} {\bibfnamefont {Stephen~H}\ \bibnamefont
  {Shenker}}, \ and\ \bibinfo {author} {\bibfnamefont {Douglas}\ \bibnamefont
  {Stanford}},\ }\bibfield  {title} {\enquote {\bibinfo {title} {A bound on
  chaos},}\ }\href@noop {} {\bibfield  {journal} {\bibinfo  {journal} {Journal
  of High Energy Physics}\ }\textbf {\bibinfo {volume} {2016}},\ \bibinfo
  {pages} {1--17} (\bibinfo {year} {2016})}\BibitemShut {NoStop}%
\bibitem [{\citenamefont {Larkin}\ and\ \citenamefont
  {Ovchinnikov}(1969)}]{larkin1969quasiclassical}%
  \BibitemOpen
  \bibfield  {author} {\bibinfo {author} {\bibfnamefont {AI}~\bibnamefont
  {Larkin}}\ and\ \bibinfo {author} {\bibfnamefont {Yu~N}\ \bibnamefont
  {Ovchinnikov}},\ }\bibfield  {title} {\enquote {\bibinfo {title}
  {Quasiclassical method in the theory of superconductivity},}\ }\href@noop {}
  {\bibfield  {journal} {\bibinfo  {journal} {Sov Phys JETP}\ }\textbf
  {\bibinfo {volume} {28}},\ \bibinfo {pages} {1200--1205} (\bibinfo {year}
  {1969})}\BibitemShut {NoStop}%
\bibitem [{\citenamefont {Hashimoto}\ \emph {et~al.}(2017)\citenamefont
  {Hashimoto}, \citenamefont {Murata},\ and\ \citenamefont
  {Yoshii}}]{hashimoto2017out}%
  \BibitemOpen
  \bibfield  {author} {\bibinfo {author} {\bibfnamefont {Koji}\ \bibnamefont
  {Hashimoto}}, \bibinfo {author} {\bibfnamefont {Keiju}\ \bibnamefont
  {Murata}}, \ and\ \bibinfo {author} {\bibfnamefont {Ryosuke}\ \bibnamefont
  {Yoshii}},\ }\bibfield  {title} {\enquote {\bibinfo {title}
  {Out-of-time-order correlators in quantum mechanics},}\ }\href@noop {}
  {\bibfield  {journal} {\bibinfo  {journal} {Journal of High Energy Physics}\
  }\textbf {\bibinfo {volume} {2017}},\ \bibinfo {pages} {1--31} (\bibinfo
  {year} {2017})}\BibitemShut {NoStop}%
\bibitem [{\citenamefont {Swingle}(2018)}]{swingle2018unscrambling}%
  \BibitemOpen
  \bibfield  {author} {\bibinfo {author} {\bibfnamefont {Brian}\ \bibnamefont
  {Swingle}},\ }\bibfield  {title} {\enquote {\bibinfo {title} {Unscrambling
  the physics of out-of-time-order correlators},}\ }\href@noop {} {\bibfield
  {journal} {\bibinfo  {journal} {Nature Physics}\ }\textbf {\bibinfo {volume}
  {14}},\ \bibinfo {pages} {988--990} (\bibinfo {year} {2018})}\BibitemShut
  {NoStop}%
\bibitem [{\citenamefont {Swingle}\ \emph {et~al.}(2016)\citenamefont
  {Swingle}, \citenamefont {Bentsen}, \citenamefont {Schleier-Smith},\ and\
  \citenamefont {Hayden}}]{swingle2016measuring}%
  \BibitemOpen
  \bibfield  {author} {\bibinfo {author} {\bibfnamefont {Brian}\ \bibnamefont
  {Swingle}}, \bibinfo {author} {\bibfnamefont {Gregory}\ \bibnamefont
  {Bentsen}}, \bibinfo {author} {\bibfnamefont {Monika}\ \bibnamefont
  {Schleier-Smith}}, \ and\ \bibinfo {author} {\bibfnamefont {Patrick}\
  \bibnamefont {Hayden}},\ }\bibfield  {title} {\enquote {\bibinfo {title}
  {Measuring the scrambling of quantum information},}\ }\href@noop {}
  {\bibfield  {journal} {\bibinfo  {journal} {Physical Review A}\ }\textbf
  {\bibinfo {volume} {94}},\ \bibinfo {pages} {040302} (\bibinfo {year}
  {2016})}\BibitemShut {NoStop}%
\bibitem [{\citenamefont {Sreeram}\ \emph {et~al.}(2021)\citenamefont
  {Sreeram}, \citenamefont {Madhok},\ and\ \citenamefont
  {Lakshminarayan}}]{sreeram2021out}%
  \BibitemOpen
  \bibfield  {author} {\bibinfo {author} {\bibfnamefont {PG}~\bibnamefont
  {Sreeram}}, \bibinfo {author} {\bibfnamefont {Vaibhav}\ \bibnamefont
  {Madhok}}, \ and\ \bibinfo {author} {\bibfnamefont {Arul}\ \bibnamefont
  {Lakshminarayan}},\ }\bibfield  {title} {\enquote {\bibinfo {title}
  {Out-of-time-ordered correlators and the loschmidt echo in the quantum kicked
  top: how low can we go?}}\ }\href@noop {} {\bibfield  {journal} {\bibinfo
  {journal} {Journal of Physics D: Applied Physics}\ }\textbf {\bibinfo
  {volume} {54}},\ \bibinfo {pages} {274004} (\bibinfo {year}
  {2021})}\BibitemShut {NoStop}%
\bibitem [{\citenamefont {Cotler}\ \emph {et~al.}(2018)\citenamefont {Cotler},
  \citenamefont {Ding},\ and\ \citenamefont {Penington}}]{cotler2018out}%
  \BibitemOpen
  \bibfield  {author} {\bibinfo {author} {\bibfnamefont {Jordan~S}\
  \bibnamefont {Cotler}}, \bibinfo {author} {\bibfnamefont {Dawei}\
  \bibnamefont {Ding}}, \ and\ \bibinfo {author} {\bibfnamefont {Geoffrey~R}\
  \bibnamefont {Penington}},\ }\bibfield  {title} {\enquote {\bibinfo {title}
  {Out-of-time-order operators and the butterfly effect},}\ }\href@noop {}
  {\bibfield  {journal} {\bibinfo  {journal} {Annals of Physics}\ }\textbf
  {\bibinfo {volume} {396}},\ \bibinfo {pages} {318--333} (\bibinfo {year}
  {2018})}\BibitemShut {NoStop}%
\bibitem [{\citenamefont {Ch{\'a}vez-Carlos}\ \emph {et~al.}(2019)\citenamefont
  {Ch{\'a}vez-Carlos}, \citenamefont {L{\'o}pez-del Carpio}, \citenamefont
  {Bastarrachea-Magnani}, \citenamefont {Str{\'a}nsk{\`y}}, \citenamefont
  {Lerma-Hern{\'a}ndez}, \citenamefont {Santos},\ and\ \citenamefont
  {Hirsch}}]{chavez2019quantum}%
  \BibitemOpen
  \bibfield  {author} {\bibinfo {author} {\bibfnamefont {Jorge}\ \bibnamefont
  {Ch{\'a}vez-Carlos}}, \bibinfo {author} {\bibfnamefont {B}~\bibnamefont
  {L{\'o}pez-del Carpio}}, \bibinfo {author} {\bibfnamefont {Miguel~A}\
  \bibnamefont {Bastarrachea-Magnani}}, \bibinfo {author} {\bibfnamefont
  {Pavel}\ \bibnamefont {Str{\'a}nsk{\`y}}}, \bibinfo {author} {\bibfnamefont
  {Sergio}\ \bibnamefont {Lerma-Hern{\'a}ndez}}, \bibinfo {author}
  {\bibfnamefont {Lea~F}\ \bibnamefont {Santos}}, \ and\ \bibinfo {author}
  {\bibfnamefont {Jorge~G}\ \bibnamefont {Hirsch}},\ }\bibfield  {title}
  {\enquote {\bibinfo {title} {Quantum and classical lyapunov exponents in
  atom-field interaction systems},}\ }\href@noop {} {\bibfield  {journal}
  {\bibinfo  {journal} {Physical review letters}\ }\textbf {\bibinfo {volume}
  {122}},\ \bibinfo {pages} {024101} (\bibinfo {year} {2019})}\BibitemShut
  {NoStop}%
\bibitem [{\citenamefont {Hashimoto}\ \emph {et~al.}(2020)\citenamefont
  {Hashimoto}, \citenamefont {Huh}, \citenamefont {Kim},\ and\ \citenamefont
  {Watanabe}}]{hashimoto2020exponential}%
  \BibitemOpen
  \bibfield  {author} {\bibinfo {author} {\bibfnamefont {Koji}\ \bibnamefont
  {Hashimoto}}, \bibinfo {author} {\bibfnamefont {Kyoung-Bum}\ \bibnamefont
  {Huh}}, \bibinfo {author} {\bibfnamefont {Keun-Young}\ \bibnamefont {Kim}}, \
  and\ \bibinfo {author} {\bibfnamefont {Ryota}\ \bibnamefont {Watanabe}},\
  }\bibfield  {title} {\enquote {\bibinfo {title} {Exponential growth of
  out-of-time-order correlator without chaos: inverted harmonic oscillator},}\
  }\href@noop {} {\bibfield  {journal} {\bibinfo  {journal} {Journal of High
  Energy Physics}\ }\textbf {\bibinfo {volume} {2020}},\ \bibinfo {pages}
  {1--25} (\bibinfo {year} {2020})}\BibitemShut {NoStop}%
\bibitem [{\citenamefont {Garc{\'\i}a-Mata}\ \emph {et~al.}(2018)\citenamefont
  {Garc{\'\i}a-Mata}, \citenamefont {Saraceno}, \citenamefont {Jalabert},
  \citenamefont {Roncaglia},\ and\ \citenamefont
  {Wisniacki}}]{garcia2018chaos}%
  \BibitemOpen
  \bibfield  {author} {\bibinfo {author} {\bibfnamefont {Ignacio}\ \bibnamefont
  {Garc{\'\i}a-Mata}}, \bibinfo {author} {\bibfnamefont {Marcos}\ \bibnamefont
  {Saraceno}}, \bibinfo {author} {\bibfnamefont {Rodolfo~A}\ \bibnamefont
  {Jalabert}}, \bibinfo {author} {\bibfnamefont {Augusto~J}\ \bibnamefont
  {Roncaglia}}, \ and\ \bibinfo {author} {\bibfnamefont {Diego~A}\ \bibnamefont
  {Wisniacki}},\ }\bibfield  {title} {\enquote {\bibinfo {title} {Chaos
  signatures in the short and long time behavior of the out-of-time ordered
  correlator},}\ }\href@noop {} {\bibfield  {journal} {\bibinfo  {journal}
  {Physical review letters}\ }\textbf {\bibinfo {volume} {121}},\ \bibinfo
  {pages} {210601} (\bibinfo {year} {2018})}\BibitemShut {NoStop}%
\bibitem [{\citenamefont {Rammensee}\ \emph {et~al.}(2018)\citenamefont
  {Rammensee}, \citenamefont {Urbina},\ and\ \citenamefont
  {Richter}}]{rammensee2018many}%
  \BibitemOpen
  \bibfield  {author} {\bibinfo {author} {\bibfnamefont {Josef}\ \bibnamefont
  {Rammensee}}, \bibinfo {author} {\bibfnamefont {Juan~Diego}\ \bibnamefont
  {Urbina}}, \ and\ \bibinfo {author} {\bibfnamefont {Klaus}\ \bibnamefont
  {Richter}},\ }\bibfield  {title} {\enquote {\bibinfo {title} {Many-body
  quantum interference and the saturation of out-of-time-order correlators},}\
  }\href@noop {} {\bibfield  {journal} {\bibinfo  {journal} {Physical Review
  Letters}\ }\textbf {\bibinfo {volume} {121}},\ \bibinfo {pages} {124101}
  (\bibinfo {year} {2018})}\BibitemShut {NoStop}%
\bibitem [{\citenamefont {Kukuljan}\ \emph {et~al.}(2017)\citenamefont
  {Kukuljan}, \citenamefont {Grozdanov},\ and\ \citenamefont
  {Prosen}}]{kukuljan2017weak}%
  \BibitemOpen
  \bibfield  {author} {\bibinfo {author} {\bibfnamefont {Ivan}\ \bibnamefont
  {Kukuljan}}, \bibinfo {author} {\bibfnamefont {Sa{\v{s}}o}\ \bibnamefont
  {Grozdanov}}, \ and\ \bibinfo {author} {\bibfnamefont {Toma{\v{z}}}\
  \bibnamefont {Prosen}},\ }\bibfield  {title} {\enquote {\bibinfo {title}
  {Weak quantum chaos},}\ }\href@noop {} {\bibfield  {journal} {\bibinfo
  {journal} {Physical Review B}\ }\textbf {\bibinfo {volume} {96}},\ \bibinfo
  {pages} {060301} (\bibinfo {year} {2017})}\BibitemShut {NoStop}%
\bibitem [{\citenamefont {Rozenbaum}\ \emph {et~al.}(2017)\citenamefont
  {Rozenbaum}, \citenamefont {Ganeshan},\ and\ \citenamefont
  {Galitski}}]{rozenbaum2017lyapunov}%
  \BibitemOpen
  \bibfield  {author} {\bibinfo {author} {\bibfnamefont {Efim~B}\ \bibnamefont
  {Rozenbaum}}, \bibinfo {author} {\bibfnamefont {Sriram}\ \bibnamefont
  {Ganeshan}}, \ and\ \bibinfo {author} {\bibfnamefont {Victor}\ \bibnamefont
  {Galitski}},\ }\bibfield  {title} {\enquote {\bibinfo {title} {Lyapunov
  exponent and out-of-time-ordered correlator’s growth rate in a chaotic
  system},}\ }\href@noop {} {\bibfield  {journal} {\bibinfo  {journal}
  {Physical review letters}\ }\textbf {\bibinfo {volume} {118}},\ \bibinfo
  {pages} {086801} (\bibinfo {year} {2017})}\BibitemShut {NoStop}%
\bibitem [{\citenamefont {Zonnios}\ \emph {et~al.}(2022)\citenamefont
  {Zonnios}, \citenamefont {Levinsen}, \citenamefont {Parish}, \citenamefont
  {Pollock},\ and\ \citenamefont {Modi}}]{modi.2022}%
  \BibitemOpen
  \bibfield  {author} {\bibinfo {author} {\bibfnamefont {Magdalini}\
  \bibnamefont {Zonnios}}, \bibinfo {author} {\bibfnamefont {Jesper}\
  \bibnamefont {Levinsen}}, \bibinfo {author} {\bibfnamefont {Meera~M.}\
  \bibnamefont {Parish}}, \bibinfo {author} {\bibfnamefont {Felix~A.}\
  \bibnamefont {Pollock}}, \ and\ \bibinfo {author} {\bibfnamefont {Kavan}\
  \bibnamefont {Modi}},\ }\bibfield  {title} {\enquote {\bibinfo {title}
  {Signatures of quantum chaos in an out-of-time-order tensor},}\ }\href
  {\doibase 10.1103/PhysRevLett.128.150601} {\bibfield  {journal} {\bibinfo
  {journal} {Phys. Rev. Lett.}\ }\textbf {\bibinfo {volume} {128}},\ \bibinfo
  {pages} {150601} (\bibinfo {year} {2022})}\BibitemShut {NoStop}%
\bibitem [{\citenamefont {Lewis-Swan}\ \emph {et~al.}(2019)\citenamefont
  {Lewis-Swan}, \citenamefont {Safavi-Naini}, \citenamefont {Bollinger},\ and\
  \citenamefont {Rey}}]{lewis2019unifying}%
  \BibitemOpen
  \bibfield  {author} {\bibinfo {author} {\bibfnamefont {RJ}~\bibnamefont
  {Lewis-Swan}}, \bibinfo {author} {\bibfnamefont {Arghavan}\ \bibnamefont
  {Safavi-Naini}}, \bibinfo {author} {\bibfnamefont {John~J}\ \bibnamefont
  {Bollinger}}, \ and\ \bibinfo {author} {\bibfnamefont {Ana~M}\ \bibnamefont
  {Rey}},\ }\bibfield  {title} {\enquote {\bibinfo {title} {Unifying
  scrambling, thermalization and entanglement through measurement of fidelity
  out-of-time-order correlators in the dicke model},}\ }\href@noop {}
  {\bibfield  {journal} {\bibinfo  {journal} {Nature communications}\ }\textbf
  {\bibinfo {volume} {10}},\ \bibinfo {pages} {1--9} (\bibinfo {year}
  {2019})}\BibitemShut {NoStop}%
\bibitem [{\citenamefont {Alba}\ and\ \citenamefont
  {Calabrese}(2019)}]{alba2019quantum}%
  \BibitemOpen
  \bibfield  {author} {\bibinfo {author} {\bibfnamefont {Vincenzo}\
  \bibnamefont {Alba}}\ and\ \bibinfo {author} {\bibfnamefont {Pasquale}\
  \bibnamefont {Calabrese}},\ }\bibfield  {title} {\enquote {\bibinfo {title}
  {Quantum information scrambling after a quantum quench},}\ }\href@noop {}
  {\bibfield  {journal} {\bibinfo  {journal} {Physical Review B}\ }\textbf
  {\bibinfo {volume} {100}},\ \bibinfo {pages} {115150} (\bibinfo {year}
  {2019})}\BibitemShut {NoStop}%
\bibitem [{\citenamefont {Styliaris}\ \emph {et~al.}(2021)\citenamefont
  {Styliaris}, \citenamefont {Anand},\ and\ \citenamefont
  {Zanardi}}]{styliaris2021information}%
  \BibitemOpen
  \bibfield  {author} {\bibinfo {author} {\bibfnamefont {Georgios}\
  \bibnamefont {Styliaris}}, \bibinfo {author} {\bibfnamefont {Namit}\
  \bibnamefont {Anand}}, \ and\ \bibinfo {author} {\bibfnamefont {Paolo}\
  \bibnamefont {Zanardi}},\ }\bibfield  {title} {\enquote {\bibinfo {title}
  {Information scrambling over bipartitions: Equilibration, entropy production,
  and typicality},}\ }\href@noop {} {\bibfield  {journal} {\bibinfo  {journal}
  {Physical Review Letters}\ }\textbf {\bibinfo {volume} {126}},\ \bibinfo
  {pages} {030601} (\bibinfo {year} {2021})}\BibitemShut {NoStop}%
\bibitem [{\citenamefont {Seshadri}\ \emph {et~al.}(2018)\citenamefont
  {Seshadri}, \citenamefont {Madhok},\ and\ \citenamefont
  {Lakshminarayan}}]{seshadri2018tripartite}%
  \BibitemOpen
  \bibfield  {author} {\bibinfo {author} {\bibfnamefont {Akshay}\ \bibnamefont
  {Seshadri}}, \bibinfo {author} {\bibfnamefont {Vaibhav}\ \bibnamefont
  {Madhok}}, \ and\ \bibinfo {author} {\bibfnamefont {Arul}\ \bibnamefont
  {Lakshminarayan}},\ }\bibfield  {title} {\enquote {\bibinfo {title}
  {Tripartite mutual information, entanglement, and scrambling in permutation
  symmetric systems with an application to quantum chaos},}\ }\href@noop {}
  {\bibfield  {journal} {\bibinfo  {journal} {Physical Review E}\ }\textbf
  {\bibinfo {volume} {98}},\ \bibinfo {pages} {052205} (\bibinfo {year}
  {2018})}\BibitemShut {NoStop}%
\bibitem [{\citenamefont {Madhok}\ \emph {et~al.}(2018)\citenamefont {Madhok},
  \citenamefont {Dogra},\ and\ \citenamefont
  {Lakshminarayan}}]{madhok2018quantum}%
  \BibitemOpen
  \bibfield  {author} {\bibinfo {author} {\bibfnamefont {Vaibhav}\ \bibnamefont
  {Madhok}}, \bibinfo {author} {\bibfnamefont {Shruti}\ \bibnamefont {Dogra}},
  \ and\ \bibinfo {author} {\bibfnamefont {Arul}\ \bibnamefont
  {Lakshminarayan}},\ }\bibfield  {title} {\enquote {\bibinfo {title} {Quantum
  correlations as probes of chaos and ergodicity},}\ }\href@noop {} {\bibfield
  {journal} {\bibinfo  {journal} {Optics Communications}\ }\textbf {\bibinfo
  {volume} {420}},\ \bibinfo {pages} {189--193} (\bibinfo {year}
  {2018})}\BibitemShut {NoStop}%
\bibitem [{\citenamefont {Madhok}\ \emph {et~al.}(2015)\citenamefont {Madhok},
  \citenamefont {Gupta}, \citenamefont {Trottier},\ and\ \citenamefont
  {Ghose}}]{madhok2015signatures}%
  \BibitemOpen
  \bibfield  {author} {\bibinfo {author} {\bibfnamefont {Vaibhav}\ \bibnamefont
  {Madhok}}, \bibinfo {author} {\bibfnamefont {Vibhu}\ \bibnamefont {Gupta}},
  \bibinfo {author} {\bibfnamefont {Denis-Alexandre}\ \bibnamefont {Trottier}},
  \ and\ \bibinfo {author} {\bibfnamefont {Shohini}\ \bibnamefont {Ghose}},\
  }\bibfield  {title} {\enquote {\bibinfo {title} {Signatures of chaos in the
  dynamics of quantum discord},}\ }\href@noop {} {\bibfield  {journal}
  {\bibinfo  {journal} {Physical Review E}\ }\textbf {\bibinfo {volume} {91}},\
  \bibinfo {pages} {032906} (\bibinfo {year} {2015})}\BibitemShut {NoStop}%
\bibitem [{\citenamefont {Latora}\ and\ \citenamefont
  {Baranger}(1999)}]{latora1999kolmogorov}%
  \BibitemOpen
  \bibfield  {author} {\bibinfo {author} {\bibfnamefont {Vito}\ \bibnamefont
  {Latora}}\ and\ \bibinfo {author} {\bibfnamefont {Michel}\ \bibnamefont
  {Baranger}},\ }\bibfield  {title} {\enquote {\bibinfo {title}
  {Kolmogorov-sinai entropy rate versus physical entropy},}\ }\href@noop {}
  {\bibfield  {journal} {\bibinfo  {journal} {Physical Review Letters}\
  }\textbf {\bibinfo {volume} {82}},\ \bibinfo {pages} {520} (\bibinfo {year}
  {1999})}\BibitemShut {NoStop}%
\bibitem [{\citenamefont {Gogolin}\ and\ \citenamefont
  {Eisert}(2016)}]{gogolin2016equilibration}%
  \BibitemOpen
  \bibfield  {author} {\bibinfo {author} {\bibfnamefont {Christian}\
  \bibnamefont {Gogolin}}\ and\ \bibinfo {author} {\bibfnamefont {Jens}\
  \bibnamefont {Eisert}},\ }\bibfield  {title} {\enquote {\bibinfo {title}
  {Equilibration, thermalisation, and the emergence of statistical mechanics in
  closed quantum systems},}\ }\href@noop {} {\bibfield  {journal} {\bibinfo
  {journal} {Reports on Progress in Physics}\ }\textbf {\bibinfo {volume}
  {79}},\ \bibinfo {pages} {056001} (\bibinfo {year} {2016})}\BibitemShut
  {NoStop}%
\bibitem [{\citenamefont {Goold}\ \emph
  {et~al.}(2016{\natexlab{a}})\citenamefont {Goold}, \citenamefont {Huber},
  \citenamefont {Riera}, \citenamefont {Del~Rio},\ and\ \citenamefont
  {Skrzypczyk}}]{goold2016role}%
  \BibitemOpen
  \bibfield  {author} {\bibinfo {author} {\bibfnamefont {John}\ \bibnamefont
  {Goold}}, \bibinfo {author} {\bibfnamefont {Marcus}\ \bibnamefont {Huber}},
  \bibinfo {author} {\bibfnamefont {Arnau}\ \bibnamefont {Riera}}, \bibinfo
  {author} {\bibfnamefont {L{\'\i}dia}\ \bibnamefont {Del~Rio}}, \ and\
  \bibinfo {author} {\bibfnamefont {Paul}\ \bibnamefont {Skrzypczyk}},\
  }\bibfield  {title} {\enquote {\bibinfo {title} {The role of quantum
  information in thermodynamics—a topical review},}\ }\href@noop {}
  {\bibfield  {journal} {\bibinfo  {journal} {Journal of Physics A:
  Mathematical and Theoretical}\ }\textbf {\bibinfo {volume} {49}},\ \bibinfo
  {pages} {143001} (\bibinfo {year} {2016}{\natexlab{a}})}\BibitemShut
  {NoStop}%
\bibitem [{\citenamefont {von Neumann}(2010)}]{von2010proof}%
  \BibitemOpen
  \bibfield  {author} {\bibinfo {author} {\bibfnamefont {John}\ \bibnamefont
  {von Neumann}},\ }\bibfield  {title} {\enquote {\bibinfo {title} {Proof of
  the ergodic theorem and the h-theorem in quantum mechanics},}\ }\href@noop {}
  {\bibfield  {journal} {\bibinfo  {journal} {The European Physical Journal H}\
  }\textbf {\bibinfo {volume} {35}},\ \bibinfo {pages} {201--237} (\bibinfo
  {year} {2010})}\BibitemShut {NoStop}%
\bibitem [{\citenamefont {Shenker}(1999)}]{She99}%
  \BibitemOpen
  \bibfield  {author} {\bibinfo {author} {\bibfnamefont {OR}~\bibnamefont
  {Shenker}},\ }\bibfield  {title} {\enquote {\bibinfo {title} {Is $- k
  \tr(\rho \ln \rho)$ the entropy in quantum mechanics},}\ }\href {\doibase
  10.1093/bjps/50.1.33} {\bibfield  {journal} {\bibinfo  {journal} {The British
  Journal for the Philosophy of Science}\ }\textbf {\bibinfo {volume} {50}},\
  \bibinfo {pages} {33--48} (\bibinfo {year} {1999})},\ \Eprint
  {http://arxiv.org/abs/http://bjps.oxfordjournals.org/content/50/1/33.full.pdf+html}
  {http://bjps.oxfordjournals.org/content/50/1/33.full.pdf+html} \BibitemShut
  {NoStop}%
\bibitem [{\citenamefont {Hemmo}\ and\ \citenamefont {Shenker}(2006)}]{HS06}%
  \BibitemOpen
  \bibfield  {author} {\bibinfo {author} {\bibfnamefont {Meir}\ \bibnamefont
  {Hemmo}}\ and\ \bibinfo {author} {\bibfnamefont {Orly}\ \bibnamefont
  {Shenker}},\ }\bibfield  {title} {\enquote {\bibinfo {title} {Von neumann’s
  entropy does not correspond to thermodynamic entropy},}\ }\href@noop {}
  {\bibfield  {journal} {\bibinfo  {journal} {Philosophy of Science}\ }\textbf
  {\bibinfo {volume} {73}},\ \bibinfo {pages} {153--174} (\bibinfo {year}
  {2006})}\BibitemShut {NoStop}%
\bibitem [{\citenamefont {Mana}\ \emph {et~al.}(2005)\citenamefont {Mana},
  \citenamefont {Maansson},\ and\ \citenamefont {Bjoerk}}]{MMB05}%
  \BibitemOpen
  \bibfield  {author} {\bibinfo {author} {\bibfnamefont {P.~G.~L.}\
  \bibnamefont {Mana}}, \bibinfo {author} {\bibfnamefont {A.}~\bibnamefont
  {Maansson}}, \ and\ \bibinfo {author} {\bibfnamefont {G.}~\bibnamefont
  {Bjoerk}},\ }\href@noop {} {\enquote {\bibinfo {title} {On
  distinguishability, orthogonality, and violations of the second law:
  contradictory assumptions, contrasting pieces of knowledge},}\ } (\bibinfo
  {year} {2005}),\ \bibinfo {note} {arXiv:quant-ph/0505229}\BibitemShut
  {NoStop}%
\bibitem [{\citenamefont {Dahlsten}\ \emph {et~al.}(2011)\citenamefont
  {Dahlsten}, \citenamefont {Renner}, \citenamefont {Rieper},\ and\
  \citenamefont {Vedral}}]{DRE+11}%
  \BibitemOpen
  \bibfield  {author} {\bibinfo {author} {\bibfnamefont {Oscar C~O}\
  \bibnamefont {Dahlsten}}, \bibinfo {author} {\bibfnamefont {Renato}\
  \bibnamefont {Renner}}, \bibinfo {author} {\bibfnamefont {Elisabeth}\
  \bibnamefont {Rieper}}, \ and\ \bibinfo {author} {\bibfnamefont {Vlatko}\
  \bibnamefont {Vedral}},\ }\bibfield  {title} {\enquote {\bibinfo {title}
  {Inadequacy of von neumann entropy for characterizing extractable work},}\
  }\href {http://stacks.iop.org/1367-2630/13/i=5/a=053015} {\bibfield
  {journal} {\bibinfo  {journal} {New Journal of Physics}\ }\textbf {\bibinfo
  {volume} {13}},\ \bibinfo {pages} {053015} (\bibinfo {year}
  {2011})}\BibitemShut {NoStop}%
\bibitem [{\citenamefont {Deville}\ and\ \citenamefont {Deville}(2013)}]{AY13}%
  \BibitemOpen
  \bibfield  {author} {\bibinfo {author} {\bibfnamefont {Alain}\ \bibnamefont
  {Deville}}\ and\ \bibinfo {author} {\bibfnamefont {Yannick}\ \bibnamefont
  {Deville}},\ }\bibfield  {title} {\enquote {\bibinfo {title} {Clarifying the
  link between von neumann and thermodynamic entropies},}\ }\href {\doibase
  10.1140/epjh/e2012-30032-0} {\bibfield  {journal} {\bibinfo  {journal} {The
  European Physical Journal H}\ }\textbf {\bibinfo {volume} {38}},\ \bibinfo
  {pages} {57--81} (\bibinfo {year} {2013})}\BibitemShut {NoStop}%
\bibitem [{\citenamefont {\ifmmode~\check{S}\else \v{S}\fi{}afr\'anek}\ \emph
  {et~al.}(2019{\natexlab{a}})\citenamefont {\ifmmode~\check{S}\else
  \v{S}\fi{}afr\'anek}, \citenamefont {Deutsch},\ and\ \citenamefont
  {Aguirre}}]{deutchquantum2019}%
  \BibitemOpen
  \bibfield  {author} {\bibinfo {author} {\bibfnamefont {Dominik}\ \bibnamefont
  {\ifmmode~\check{S}\else \v{S}\fi{}afr\'anek}}, \bibinfo {author}
  {\bibfnamefont {J.~M.}\ \bibnamefont {Deutsch}}, \ and\ \bibinfo {author}
  {\bibfnamefont {Anthony}\ \bibnamefont {Aguirre}},\ }\bibfield  {title}
  {\enquote {\bibinfo {title} {Quantum coarse-grained entropy and
  thermodynamics},}\ }\href {\doibase 10.1103/PhysRevA.99.010101} {\bibfield
  {journal} {\bibinfo  {journal} {Phys. Rev. A}\ }\textbf {\bibinfo {volume}
  {99}},\ \bibinfo {pages} {010101} (\bibinfo {year}
  {2019}{\natexlab{a}})}\BibitemShut {NoStop}%
\bibitem [{\citenamefont {\ifmmode~\check{S}\else \v{S}\fi{}afr\'anek}\ \emph
  {et~al.}(2019{\natexlab{b}})\citenamefont {\ifmmode~\check{S}\else
  \v{S}\fi{}afr\'anek}, \citenamefont {Deutsch},\ and\ \citenamefont
  {Aguirre}}]{quantum2020deautch}%
  \BibitemOpen
  \bibfield  {author} {\bibinfo {author} {\bibfnamefont {Dominik}\ \bibnamefont
  {\ifmmode~\check{S}\else \v{S}\fi{}afr\'anek}}, \bibinfo {author}
  {\bibfnamefont {J.~M.}\ \bibnamefont {Deutsch}}, \ and\ \bibinfo {author}
  {\bibfnamefont {Anthony}\ \bibnamefont {Aguirre}},\ }\bibfield  {title}
  {\enquote {\bibinfo {title} {Quantum coarse-grained entropy and
  thermalization in closed systems},}\ }\href {\doibase
  10.1103/PhysRevA.99.012103} {\bibfield  {journal} {\bibinfo  {journal} {Phys.
  Rev. A}\ }\textbf {\bibinfo {volume} {99}},\ \bibinfo {pages} {012103}
  (\bibinfo {year} {2019}{\natexlab{b}})}\BibitemShut {NoStop}%
\bibitem [{\citenamefont {Gibbs}(1902)}]{gibbs1902elementary}%
  \BibitemOpen
  \bibfield  {author} {\bibinfo {author} {\bibfnamefont {Josiah~Willard}\
  \bibnamefont {Gibbs}},\ }\bibfield  {title} {\enquote {\bibinfo {title}
  {Elementary principles in statistical mechanics: developed with especial
  reference to the rational foundations of thermodynamics},}\ }\href@noop {} {\
   (\bibinfo {year} {1902})}\BibitemShut {NoStop}%
\bibitem [{\citenamefont {Ehrenfest}\ and\ \citenamefont
  {Ehrenfest}(1990)}]{ehrenfest1990conceptual}%
  \BibitemOpen
  \bibfield  {author} {\bibinfo {author} {\bibfnamefont {Paul}\ \bibnamefont
  {Ehrenfest}}\ and\ \bibinfo {author} {\bibfnamefont {Tatiana}\ \bibnamefont
  {Ehrenfest}},\ }\bibfield  {title} {\enquote {\bibinfo {title} {The
  conceptual foundations of the statistical approach in mechanics},}\
  }\href@noop {} {\  (\bibinfo {year} {1990})}\BibitemShut {NoStop}%
\bibitem [{\citenamefont {Caves}()}]{Caves_notes}%
  \BibitemOpen
  \bibfield  {author} {\bibinfo {author} {\bibfnamefont {Carlton~M.}\
  \bibnamefont {Caves}},\ }\href@noop {} {\enquote {\bibinfo {title} {Resource
  material for promoting the bayesian view of everything},}\ }\bibinfo {note}
  {Personal notes}\BibitemShut {NoStop}%
\bibitem [{\citenamefont {{\v{S}}afr{\'a}nek}\ \emph
  {et~al.}(2021)\citenamefont {{\v{S}}afr{\'a}nek}, \citenamefont {Aguirre},
  \citenamefont {Schindler},\ and\ \citenamefont
  {Deutsch}}]{vsafranek2021brief}%
  \BibitemOpen
  \bibfield  {author} {\bibinfo {author} {\bibfnamefont {Dominik}\ \bibnamefont
  {{\v{S}}afr{\'a}nek}}, \bibinfo {author} {\bibfnamefont {Anthony}\
  \bibnamefont {Aguirre}}, \bibinfo {author} {\bibfnamefont {Joseph}\
  \bibnamefont {Schindler}}, \ and\ \bibinfo {author} {\bibfnamefont
  {JM}~\bibnamefont {Deutsch}},\ }\bibfield  {title} {\enquote {\bibinfo
  {title} {A brief introduction to observational entropy},}\ }\href@noop {}
  {\bibfield  {journal} {\bibinfo  {journal} {Foundations of Physics}\ }\textbf
  {\bibinfo {volume} {51}},\ \bibinfo {pages} {1--20} (\bibinfo {year}
  {2021})}\BibitemShut {NoStop}%
\bibitem [{\citenamefont {Strasberg}\ and\ \citenamefont
  {Winter}(2021)}]{strasberg2021second}%
  \BibitemOpen
  \bibfield  {author} {\bibinfo {author} {\bibfnamefont {Philipp}\ \bibnamefont
  {Strasberg}}\ and\ \bibinfo {author} {\bibfnamefont {Andreas}\ \bibnamefont
  {Winter}},\ }\bibfield  {title} {\enquote {\bibinfo {title} {First and second
  law of quantum thermodynamics: a consistent derivation based on a microscopic
  definition of entropy},}\ }\href@noop {} {\bibfield  {journal} {\bibinfo
  {journal} {PRX Quantum}\ }\textbf {\bibinfo {volume} {2}},\ \bibinfo {pages}
  {030202} (\bibinfo {year} {2021})}\BibitemShut {NoStop}%
\bibitem [{\citenamefont {{\v{S}}afr{\'a}nek}\ \emph
  {et~al.}(2020)\citenamefont {{\v{S}}afr{\'a}nek}, \citenamefont {Aguirre},\
  and\ \citenamefont {Deutsch}}]{vsafranek2020classical}%
  \BibitemOpen
  \bibfield  {author} {\bibinfo {author} {\bibfnamefont {Dominik}\ \bibnamefont
  {{\v{S}}afr{\'a}nek}}, \bibinfo {author} {\bibfnamefont {Anthony}\
  \bibnamefont {Aguirre}}, \ and\ \bibinfo {author} {\bibfnamefont
  {JM}~\bibnamefont {Deutsch}},\ }\bibfield  {title} {\enquote {\bibinfo
  {title} {Classical dynamical coarse-grained entropy and comparison with the
  quantum version},}\ }\href@noop {} {\bibfield  {journal} {\bibinfo  {journal}
  {Physical Review E}\ }\textbf {\bibinfo {volume} {102}},\ \bibinfo {pages}
  {032106} (\bibinfo {year} {2020})}\BibitemShut {NoStop}%
\bibitem [{\citenamefont {Buscemi}\ \emph
  {et~al.}(2022{\natexlab{a}})\citenamefont {Buscemi}, \citenamefont
  {Schindler},\ and\ \citenamefont {Šafránek}}]{Buscemi_22}%
  \BibitemOpen
  \bibfield  {author} {\bibinfo {author} {\bibfnamefont {Francesco}\
  \bibnamefont {Buscemi}}, \bibinfo {author} {\bibfnamefont {Joseph}\
  \bibnamefont {Schindler}}, \ and\ \bibinfo {author} {\bibfnamefont {Dominik}\
  \bibnamefont {Šafránek}},\ }\href {\doibase 10.48550/ARXIV.2209.03803}
  {\enquote {\bibinfo {title} {Observational entropy, coarse quantum states,
  and petz recovery: information-theoretic properties and bounds},}\ }
  (\bibinfo {year} {2022}{\natexlab{a}})\BibitemShut {NoStop}%
\bibitem [{\citenamefont {Schindler}\ \emph {et~al.}(2020)\citenamefont
  {Schindler}, \citenamefont {{\v{S}}afr{\'a}nek},\ and\ \citenamefont
  {Aguirre}}]{schindler2020quantum}%
  \BibitemOpen
  \bibfield  {author} {\bibinfo {author} {\bibfnamefont {Joseph}\ \bibnamefont
  {Schindler}}, \bibinfo {author} {\bibfnamefont {Dominik}\ \bibnamefont
  {{\v{S}}afr{\'a}nek}}, \ and\ \bibinfo {author} {\bibfnamefont {Anthony}\
  \bibnamefont {Aguirre}},\ }\bibfield  {title} {\enquote {\bibinfo {title}
  {Quantum correlation entropy},}\ }\href@noop {} {\bibfield  {journal}
  {\bibinfo  {journal} {Physical Review A}\ }\textbf {\bibinfo {volume}
  {102}},\ \bibinfo {pages} {052407} (\bibinfo {year} {2020})}\BibitemShut
  {NoStop}%
\bibitem [{\citenamefont {Carr}\ \emph {et~al.}(2009)\citenamefont {Carr},
  \citenamefont {DeMille}, \citenamefont {Krems},\ and\ \citenamefont
  {Ye}}]{carr2009cold}%
  \BibitemOpen
  \bibfield  {author} {\bibinfo {author} {\bibfnamefont {Lincoln~D}\
  \bibnamefont {Carr}}, \bibinfo {author} {\bibfnamefont {David}\ \bibnamefont
  {DeMille}}, \bibinfo {author} {\bibfnamefont {Roman~V}\ \bibnamefont
  {Krems}}, \ and\ \bibinfo {author} {\bibfnamefont {Jun}\ \bibnamefont {Ye}},\
  }\bibfield  {title} {\enquote {\bibinfo {title} {Cold and ultracold
  molecules: science, technology and applications},}\ }\href@noop {} {\bibfield
   {journal} {\bibinfo  {journal} {New Journal of Physics}\ }\textbf {\bibinfo
  {volume} {11}},\ \bibinfo {pages} {055049} (\bibinfo {year}
  {2009})}\BibitemShut {NoStop}%
\bibitem [{\citenamefont {Safronova}\ \emph {et~al.}(2018)\citenamefont
  {Safronova}, \citenamefont {Budker}, \citenamefont {DeMille}, \citenamefont
  {Kimball}, \citenamefont {Derevianko},\ and\ \citenamefont
  {Clark}}]{safronova2018search}%
  \BibitemOpen
  \bibfield  {author} {\bibinfo {author} {\bibfnamefont {MS}~\bibnamefont
  {Safronova}}, \bibinfo {author} {\bibfnamefont {D}~\bibnamefont {Budker}},
  \bibinfo {author} {\bibfnamefont {D}~\bibnamefont {DeMille}}, \bibinfo
  {author} {\bibfnamefont {Derek F~Jackson}\ \bibnamefont {Kimball}}, \bibinfo
  {author} {\bibfnamefont {A}~\bibnamefont {Derevianko}}, \ and\ \bibinfo
  {author} {\bibfnamefont {Charles~W}\ \bibnamefont {Clark}},\ }\bibfield
  {title} {\enquote {\bibinfo {title} {Search for new physics with atoms and
  molecules},}\ }\href@noop {} {\bibfield  {journal} {\bibinfo  {journal}
  {Reviews of Modern Physics}\ }\textbf {\bibinfo {volume} {90}},\ \bibinfo
  {pages} {025008} (\bibinfo {year} {2018})}\BibitemShut {NoStop}%
\bibitem [{\citenamefont {Atat{\"u}re}\ \emph {et~al.}(2018)\citenamefont
  {Atat{\"u}re}, \citenamefont {Englund}, \citenamefont {Vamivakas},
  \citenamefont {Lee},\ and\ \citenamefont {Wrachtrup}}]{atature2018material}%
  \BibitemOpen
  \bibfield  {author} {\bibinfo {author} {\bibfnamefont {Mete}\ \bibnamefont
  {Atat{\"u}re}}, \bibinfo {author} {\bibfnamefont {Dirk}\ \bibnamefont
  {Englund}}, \bibinfo {author} {\bibfnamefont {Nick}\ \bibnamefont
  {Vamivakas}}, \bibinfo {author} {\bibfnamefont {Sang-Yun}\ \bibnamefont
  {Lee}}, \ and\ \bibinfo {author} {\bibfnamefont {Joerg}\ \bibnamefont
  {Wrachtrup}},\ }\bibfield  {title} {\enquote {\bibinfo {title} {Material
  platforms for spin-based photonic quantum technologies},}\ }\href@noop {}
  {\bibfield  {journal} {\bibinfo  {journal} {Nature Reviews Materials}\
  }\textbf {\bibinfo {volume} {3}},\ \bibinfo {pages} {38--51} (\bibinfo {year}
  {2018})}\BibitemShut {NoStop}%
\bibitem [{\citenamefont {Vandersypen}\ and\ \citenamefont
  {Eriksson}(2019)}]{vandersypen2019semiconductor}%
  \BibitemOpen
  \bibfield  {author} {\bibinfo {author} {\bibfnamefont {Lieven~MK}\
  \bibnamefont {Vandersypen}}\ and\ \bibinfo {author} {\bibfnamefont {Mark~A}\
  \bibnamefont {Eriksson}},\ }\bibfield  {title} {\enquote {\bibinfo {title}
  {Quantum computing with semiconductor spins},}\ }\href@noop {} {\bibfield
  {journal} {\bibinfo  {journal} {Physics Today}\ }\textbf {\bibinfo {volume}
  {72}},\ \bibinfo {pages} {8--38} (\bibinfo {year} {2019})}\BibitemShut
  {NoStop}%
\bibitem [{\citenamefont {Hensgens}\ \emph {et~al.}(2017)\citenamefont
  {Hensgens}, \citenamefont {Fujita}, \citenamefont {Janssen}, \citenamefont
  {Li}, \citenamefont {Van~Diepen}, \citenamefont {Reichl}, \citenamefont
  {Wegscheider}, \citenamefont {Das~Sarma},\ and\ \citenamefont
  {Vandersypen}}]{hensgens2017quantum}%
  \BibitemOpen
  \bibfield  {author} {\bibinfo {author} {\bibfnamefont {Toivo}\ \bibnamefont
  {Hensgens}}, \bibinfo {author} {\bibfnamefont {Takafumi}\ \bibnamefont
  {Fujita}}, \bibinfo {author} {\bibfnamefont {Laurens}\ \bibnamefont
  {Janssen}}, \bibinfo {author} {\bibfnamefont {Xiao}\ \bibnamefont {Li}},
  \bibinfo {author} {\bibfnamefont {CJ}~\bibnamefont {Van~Diepen}}, \bibinfo
  {author} {\bibfnamefont {Christian}\ \bibnamefont {Reichl}}, \bibinfo
  {author} {\bibfnamefont {Werner}\ \bibnamefont {Wegscheider}}, \bibinfo
  {author} {\bibfnamefont {Sankar}\ \bibnamefont {Das~Sarma}}, \ and\ \bibinfo
  {author} {\bibfnamefont {Lieven~MK}\ \bibnamefont {Vandersypen}},\ }\bibfield
   {title} {\enquote {\bibinfo {title} {Quantum simulation of a fermi--hubbard
  model using a semiconductor quantum dot array},}\ }\href@noop {} {\bibfield
  {journal} {\bibinfo  {journal} {Nature}\ }\textbf {\bibinfo {volume} {548}},\
  \bibinfo {pages} {70--73} (\bibinfo {year} {2017})}\BibitemShut {NoStop}%
\bibitem [{\citenamefont {Hartmann}\ \emph {et~al.}(2008)\citenamefont
  {Hartmann}, \citenamefont {Brandao},\ and\ \citenamefont
  {Plenio}}]{hartmann2008quantum}%
  \BibitemOpen
  \bibfield  {author} {\bibinfo {author} {\bibfnamefont {Michael~J}\
  \bibnamefont {Hartmann}}, \bibinfo {author} {\bibfnamefont {Fernando~GSL}\
  \bibnamefont {Brandao}}, \ and\ \bibinfo {author} {\bibfnamefont {Martin~B}\
  \bibnamefont {Plenio}},\ }\bibfield  {title} {\enquote {\bibinfo {title}
  {Quantum many-body phenomena in coupled cavity arrays},}\ }\href@noop {}
  {\bibfield  {journal} {\bibinfo  {journal} {Laser \& Photonics Reviews}\
  }\textbf {\bibinfo {volume} {2}},\ \bibinfo {pages} {527--556} (\bibinfo
  {year} {2008})}\BibitemShut {NoStop}%
\bibitem [{\citenamefont {Vaidya}\ \emph {et~al.}(2018)\citenamefont {Vaidya},
  \citenamefont {Guo}, \citenamefont {Kroeze}, \citenamefont {Ballantine},
  \citenamefont {Koll{\'a}r}, \citenamefont {Keeling},\ and\ \citenamefont
  {Lev}}]{vaidya2018tunable}%
  \BibitemOpen
  \bibfield  {author} {\bibinfo {author} {\bibfnamefont {Varun~D}\ \bibnamefont
  {Vaidya}}, \bibinfo {author} {\bibfnamefont {Yudan}\ \bibnamefont {Guo}},
  \bibinfo {author} {\bibfnamefont {Ronen~M}\ \bibnamefont {Kroeze}}, \bibinfo
  {author} {\bibfnamefont {Kyle~E}\ \bibnamefont {Ballantine}}, \bibinfo
  {author} {\bibfnamefont {Alicia~J}\ \bibnamefont {Koll{\'a}r}}, \bibinfo
  {author} {\bibfnamefont {Jonathan}\ \bibnamefont {Keeling}}, \ and\ \bibinfo
  {author} {\bibfnamefont {Benjamin~L}\ \bibnamefont {Lev}},\ }\bibfield
  {title} {\enquote {\bibinfo {title} {Tunable-range, photon-mediated atomic
  interactions in multimode cavity qed},}\ }\href@noop {} {\bibfield  {journal}
  {\bibinfo  {journal} {Physical Review X}\ }\textbf {\bibinfo {volume} {8}},\
  \bibinfo {pages} {011002} (\bibinfo {year} {2018})}\BibitemShut {NoStop}%
\bibitem [{\citenamefont {Norcia}\ \emph {et~al.}(2018)\citenamefont {Norcia},
  \citenamefont {Lewis-Swan}, \citenamefont {Cline}, \citenamefont {Zhu},
  \citenamefont {Rey},\ and\ \citenamefont {Thompson}}]{norcia2018cavity}%
  \BibitemOpen
  \bibfield  {author} {\bibinfo {author} {\bibfnamefont {Matthew~A}\
  \bibnamefont {Norcia}}, \bibinfo {author} {\bibfnamefont {Robert~J}\
  \bibnamefont {Lewis-Swan}}, \bibinfo {author} {\bibfnamefont {Julia~RK}\
  \bibnamefont {Cline}}, \bibinfo {author} {\bibfnamefont {Bihui}\ \bibnamefont
  {Zhu}}, \bibinfo {author} {\bibfnamefont {Ana~M}\ \bibnamefont {Rey}}, \ and\
  \bibinfo {author} {\bibfnamefont {James~K}\ \bibnamefont {Thompson}},\
  }\bibfield  {title} {\enquote {\bibinfo {title} {Cavity-mediated collective
  spin-exchange interactions in a strontium superradiant laser},}\ }\href@noop
  {} {\bibfield  {journal} {\bibinfo  {journal} {Science}\ }\textbf {\bibinfo
  {volume} {361}},\ \bibinfo {pages} {259--262} (\bibinfo {year}
  {2018})}\BibitemShut {NoStop}%
\bibitem [{\citenamefont {Davis}\ \emph {et~al.}(2019)\citenamefont {Davis},
  \citenamefont {Bentsen}, \citenamefont {Homeier}, \citenamefont {Li},\ and\
  \citenamefont {Schleier-Smith}}]{davis2019photon}%
  \BibitemOpen
  \bibfield  {author} {\bibinfo {author} {\bibfnamefont {Emily~J}\ \bibnamefont
  {Davis}}, \bibinfo {author} {\bibfnamefont {Gregory}\ \bibnamefont
  {Bentsen}}, \bibinfo {author} {\bibfnamefont {Lukas}\ \bibnamefont
  {Homeier}}, \bibinfo {author} {\bibfnamefont {Tracy}\ \bibnamefont {Li}}, \
  and\ \bibinfo {author} {\bibfnamefont {Monika~H}\ \bibnamefont
  {Schleier-Smith}},\ }\bibfield  {title} {\enquote {\bibinfo {title}
  {Photon-mediated spin-exchange dynamics of spin-1 atoms},}\ }\href@noop {}
  {\bibfield  {journal} {\bibinfo  {journal} {Physical review letters}\
  }\textbf {\bibinfo {volume} {122}},\ \bibinfo {pages} {010405} (\bibinfo
  {year} {2019})}\BibitemShut {NoStop}%
\bibitem [{\citenamefont {Wang}\ \emph {et~al.}(2015)\citenamefont {Wang},
  \citenamefont {Cai}, \citenamefont {Yuan}, \citenamefont {Zhu},\ and\
  \citenamefont {Liu}}]{wang2015topological}%
  \BibitemOpen
  \bibfield  {author} {\bibinfo {author} {\bibfnamefont {Da-Wei}\ \bibnamefont
  {Wang}}, \bibinfo {author} {\bibfnamefont {Han}\ \bibnamefont {Cai}},
  \bibinfo {author} {\bibfnamefont {Luqi}\ \bibnamefont {Yuan}}, \bibinfo
  {author} {\bibfnamefont {Shi-Yao}\ \bibnamefont {Zhu}}, \ and\ \bibinfo
  {author} {\bibfnamefont {Ren-Bao}\ \bibnamefont {Liu}},\ }\bibfield  {title}
  {\enquote {\bibinfo {title} {Topological phase transitions in superradiance
  lattices},}\ }\href@noop {} {\bibfield  {journal} {\bibinfo  {journal}
  {Optica}\ }\textbf {\bibinfo {volume} {2}},\ \bibinfo {pages} {712--715}
  (\bibinfo {year} {2015})}\BibitemShut {NoStop}%
\bibitem [{\citenamefont {Noh}\ and\ \citenamefont
  {Angelakis}(2016)}]{noh2016quantum}%
  \BibitemOpen
  \bibfield  {author} {\bibinfo {author} {\bibfnamefont {Changsuk}\
  \bibnamefont {Noh}}\ and\ \bibinfo {author} {\bibfnamefont {Dimitris~G}\
  \bibnamefont {Angelakis}},\ }\bibfield  {title} {\enquote {\bibinfo {title}
  {Quantum simulations and many-body physics with light},}\ }\href@noop {}
  {\bibfield  {journal} {\bibinfo  {journal} {Reports on Progress in Physics}\
  }\textbf {\bibinfo {volume} {80}},\ \bibinfo {pages} {016401} (\bibinfo
  {year} {2016})}\BibitemShut {NoStop}%
\bibitem [{\citenamefont {Hartmann}(2016)}]{hartmann2016quantum}%
  \BibitemOpen
  \bibfield  {author} {\bibinfo {author} {\bibfnamefont {Michael~J}\
  \bibnamefont {Hartmann}},\ }\bibfield  {title} {\enquote {\bibinfo {title}
  {Quantum simulation with interacting photons},}\ }\href@noop {} {\bibfield
  {journal} {\bibinfo  {journal} {Journal of Optics}\ }\textbf {\bibinfo
  {volume} {18}},\ \bibinfo {pages} {104005} (\bibinfo {year}
  {2016})}\BibitemShut {NoStop}%
\bibitem [{\citenamefont {Kjaergaard}\ \emph {et~al.}(2020)\citenamefont
  {Kjaergaard}, \citenamefont {Schwartz}, \citenamefont {Braum{\"u}ller},
  \citenamefont {Krantz}, \citenamefont {Wang}, \citenamefont {Gustavsson},\
  and\ \citenamefont {Oliver}}]{kjaergaard2020superconducting}%
  \BibitemOpen
  \bibfield  {author} {\bibinfo {author} {\bibfnamefont {Morten}\ \bibnamefont
  {Kjaergaard}}, \bibinfo {author} {\bibfnamefont {Mollie~E}\ \bibnamefont
  {Schwartz}}, \bibinfo {author} {\bibfnamefont {Jochen}\ \bibnamefont
  {Braum{\"u}ller}}, \bibinfo {author} {\bibfnamefont {Philip}\ \bibnamefont
  {Krantz}}, \bibinfo {author} {\bibfnamefont {Joel I-J}\ \bibnamefont {Wang}},
  \bibinfo {author} {\bibfnamefont {Simon}\ \bibnamefont {Gustavsson}}, \ and\
  \bibinfo {author} {\bibfnamefont {William~D}\ \bibnamefont {Oliver}},\
  }\bibfield  {title} {\enquote {\bibinfo {title} {Superconducting qubits:
  Current state of play},}\ }\href@noop {} {\bibfield  {journal} {\bibinfo
  {journal} {Annual Review of Condensed Matter Physics}\ }\textbf {\bibinfo
  {volume} {11}},\ \bibinfo {pages} {369--395} (\bibinfo {year}
  {2020})}\BibitemShut {NoStop}%
\bibitem [{\citenamefont {Ganzhorn}\ \emph {et~al.}(2019)\citenamefont
  {Ganzhorn}, \citenamefont {Egger}, \citenamefont {Barkoutsos}, \citenamefont
  {Ollitrault}, \citenamefont {Salis}, \citenamefont {Moll}, \citenamefont
  {Roth}, \citenamefont {Fuhrer}, \citenamefont {Mueller}, \citenamefont
  {Woerner} \emph {et~al.}}]{ganzhorn2019gate}%
  \BibitemOpen
  \bibfield  {author} {\bibinfo {author} {\bibfnamefont {Marc}\ \bibnamefont
  {Ganzhorn}}, \bibinfo {author} {\bibfnamefont {Daniel~J}\ \bibnamefont
  {Egger}}, \bibinfo {author} {\bibfnamefont {Panagiotis}\ \bibnamefont
  {Barkoutsos}}, \bibinfo {author} {\bibfnamefont {Pauline}\ \bibnamefont
  {Ollitrault}}, \bibinfo {author} {\bibfnamefont {Gian}\ \bibnamefont
  {Salis}}, \bibinfo {author} {\bibfnamefont {Nikolaj}\ \bibnamefont {Moll}},
  \bibinfo {author} {\bibfnamefont {Marco}\ \bibnamefont {Roth}}, \bibinfo
  {author} {\bibfnamefont {A}~\bibnamefont {Fuhrer}}, \bibinfo {author}
  {\bibfnamefont {P}~\bibnamefont {Mueller}}, \bibinfo {author} {\bibfnamefont
  {Stefan}\ \bibnamefont {Woerner}},  \emph {et~al.},\ }\bibfield  {title}
  {\enquote {\bibinfo {title} {Gate-efficient simulation of molecular
  eigenstates on a quantum computer},}\ }\href@noop {} {\bibfield  {journal}
  {\bibinfo  {journal} {Physical Review Applied}\ }\textbf {\bibinfo {volume}
  {11}},\ \bibinfo {pages} {044092} (\bibinfo {year} {2019})}\BibitemShut
  {NoStop}%
\bibitem [{\citenamefont {Kandala}\ \emph {et~al.}(2017)\citenamefont
  {Kandala}, \citenamefont {Mezzacapo}, \citenamefont {Temme}, \citenamefont
  {Takita}, \citenamefont {Brink}, \citenamefont {Chow},\ and\ \citenamefont
  {Gambetta}}]{kandala2017hardware}%
  \BibitemOpen
  \bibfield  {author} {\bibinfo {author} {\bibfnamefont {Abhinav}\ \bibnamefont
  {Kandala}}, \bibinfo {author} {\bibfnamefont {Antonio}\ \bibnamefont
  {Mezzacapo}}, \bibinfo {author} {\bibfnamefont {Kristan}\ \bibnamefont
  {Temme}}, \bibinfo {author} {\bibfnamefont {Maika}\ \bibnamefont {Takita}},
  \bibinfo {author} {\bibfnamefont {Markus}\ \bibnamefont {Brink}}, \bibinfo
  {author} {\bibfnamefont {Jerry~M}\ \bibnamefont {Chow}}, \ and\ \bibinfo
  {author} {\bibfnamefont {Jay~M}\ \bibnamefont {Gambetta}},\ }\bibfield
  {title} {\enquote {\bibinfo {title} {Hardware-efficient variational quantum
  eigensolver for small molecules and quantum magnets},}\ }\href@noop {}
  {\bibfield  {journal} {\bibinfo  {journal} {Nature}\ }\textbf {\bibinfo
  {volume} {549}},\ \bibinfo {pages} {242--246} (\bibinfo {year}
  {2017})}\BibitemShut {NoStop}%
\bibitem [{\citenamefont {Hempel}\ \emph {et~al.}(2018)\citenamefont {Hempel},
  \citenamefont {Maier}, \citenamefont {Romero}, \citenamefont {McClean},
  \citenamefont {Monz}, \citenamefont {Shen}, \citenamefont {Jurcevic},
  \citenamefont {Lanyon}, \citenamefont {Love}, \citenamefont {Babbush} \emph
  {et~al.}}]{hempel2018quantum}%
  \BibitemOpen
  \bibfield  {author} {\bibinfo {author} {\bibfnamefont {Cornelius}\
  \bibnamefont {Hempel}}, \bibinfo {author} {\bibfnamefont {Christine}\
  \bibnamefont {Maier}}, \bibinfo {author} {\bibfnamefont {Jonathan}\
  \bibnamefont {Romero}}, \bibinfo {author} {\bibfnamefont {Jarrod}\
  \bibnamefont {McClean}}, \bibinfo {author} {\bibfnamefont {Thomas}\
  \bibnamefont {Monz}}, \bibinfo {author} {\bibfnamefont {Heng}\ \bibnamefont
  {Shen}}, \bibinfo {author} {\bibfnamefont {Petar}\ \bibnamefont {Jurcevic}},
  \bibinfo {author} {\bibfnamefont {Ben~P}\ \bibnamefont {Lanyon}}, \bibinfo
  {author} {\bibfnamefont {Peter}\ \bibnamefont {Love}}, \bibinfo {author}
  {\bibfnamefont {Ryan}\ \bibnamefont {Babbush}},  \emph {et~al.},\ }\bibfield
  {title} {\enquote {\bibinfo {title} {Quantum chemistry calculations on a
  trapped-ion quantum simulator},}\ }\href@noop {} {\bibfield  {journal}
  {\bibinfo  {journal} {Physical Review X}\ }\textbf {\bibinfo {volume} {8}},\
  \bibinfo {pages} {031022} (\bibinfo {year} {2018})}\BibitemShut {NoStop}%
\bibitem [{\citenamefont {Nam}\ \emph {et~al.}(2020)\citenamefont {Nam},
  \citenamefont {Chen}, \citenamefont {Pisenti}, \citenamefont {Wright},
  \citenamefont {Delaney}, \citenamefont {Maslov}, \citenamefont {Brown},
  \citenamefont {Allen}, \citenamefont {Amini}, \citenamefont {Apisdorf} \emph
  {et~al.}}]{nam2020ground}%
  \BibitemOpen
  \bibfield  {author} {\bibinfo {author} {\bibfnamefont {Yunseong}\
  \bibnamefont {Nam}}, \bibinfo {author} {\bibfnamefont {Jwo-Sy}\ \bibnamefont
  {Chen}}, \bibinfo {author} {\bibfnamefont {Neal~C}\ \bibnamefont {Pisenti}},
  \bibinfo {author} {\bibfnamefont {Kenneth}\ \bibnamefont {Wright}}, \bibinfo
  {author} {\bibfnamefont {Conor}\ \bibnamefont {Delaney}}, \bibinfo {author}
  {\bibfnamefont {Dmitri}\ \bibnamefont {Maslov}}, \bibinfo {author}
  {\bibfnamefont {Kenneth~R}\ \bibnamefont {Brown}}, \bibinfo {author}
  {\bibfnamefont {Stewart}\ \bibnamefont {Allen}}, \bibinfo {author}
  {\bibfnamefont {Jason~M}\ \bibnamefont {Amini}}, \bibinfo {author}
  {\bibfnamefont {Joel}\ \bibnamefont {Apisdorf}},  \emph {et~al.},\ }\bibfield
   {title} {\enquote {\bibinfo {title} {Ground-state energy estimation of the
  water molecule on a trapped-ion quantum computer},}\ }\href@noop {}
  {\bibfield  {journal} {\bibinfo  {journal} {npj Quantum Information}\
  }\textbf {\bibinfo {volume} {6}},\ \bibinfo {pages} {1--6} (\bibinfo {year}
  {2020})}\BibitemShut {NoStop}%
\bibitem [{\citenamefont {Britton}\ \emph {et~al.}(2012)\citenamefont
  {Britton}, \citenamefont {Sawyer}, \citenamefont {Keith}, \citenamefont
  {Wang}, \citenamefont {Freericks}, \citenamefont {Uys}, \citenamefont
  {Biercuk},\ and\ \citenamefont {Bollinger}}]{britton2012engineered}%
  \BibitemOpen
  \bibfield  {author} {\bibinfo {author} {\bibfnamefont {Joseph~W}\
  \bibnamefont {Britton}}, \bibinfo {author} {\bibfnamefont {Brian~C}\
  \bibnamefont {Sawyer}}, \bibinfo {author} {\bibfnamefont {Adam~C}\
  \bibnamefont {Keith}}, \bibinfo {author} {\bibfnamefont {C-C~Joseph}\
  \bibnamefont {Wang}}, \bibinfo {author} {\bibfnamefont {James~K}\
  \bibnamefont {Freericks}}, \bibinfo {author} {\bibfnamefont {Hermann}\
  \bibnamefont {Uys}}, \bibinfo {author} {\bibfnamefont {Michael~J}\
  \bibnamefont {Biercuk}}, \ and\ \bibinfo {author} {\bibfnamefont {John~J}\
  \bibnamefont {Bollinger}},\ }\bibfield  {title} {\enquote {\bibinfo {title}
  {Engineered two-dimensional ising interactions in a trapped-ion quantum
  simulator with hundreds of spins},}\ }\href@noop {} {\bibfield  {journal}
  {\bibinfo  {journal} {Nature}\ }\textbf {\bibinfo {volume} {484}},\ \bibinfo
  {pages} {489--492} (\bibinfo {year} {2012})}\BibitemShut {NoStop}%
\bibitem [{\citenamefont {Bohnet}\ \emph {et~al.}(2016)\citenamefont {Bohnet},
  \citenamefont {Sawyer}, \citenamefont {Britton}, \citenamefont {Wall},
  \citenamefont {Rey}, \citenamefont {Foss-Feig},\ and\ \citenamefont
  {Bollinger}}]{bohnet2016quantum}%
  \BibitemOpen
  \bibfield  {author} {\bibinfo {author} {\bibfnamefont {Justin~G}\
  \bibnamefont {Bohnet}}, \bibinfo {author} {\bibfnamefont {Brian~C}\
  \bibnamefont {Sawyer}}, \bibinfo {author} {\bibfnamefont {Joseph~W}\
  \bibnamefont {Britton}}, \bibinfo {author} {\bibfnamefont {Michael~L}\
  \bibnamefont {Wall}}, \bibinfo {author} {\bibfnamefont {Ana~Maria}\
  \bibnamefont {Rey}}, \bibinfo {author} {\bibfnamefont {Michael}\ \bibnamefont
  {Foss-Feig}}, \ and\ \bibinfo {author} {\bibfnamefont {John~J}\ \bibnamefont
  {Bollinger}},\ }\bibfield  {title} {\enquote {\bibinfo {title} {Quantum spin
  dynamics and entanglement generation with hundreds of trapped ions},}\
  }\href@noop {} {\bibfield  {journal} {\bibinfo  {journal} {Science}\ }\textbf
  {\bibinfo {volume} {352}},\ \bibinfo {pages} {1297--1301} (\bibinfo {year}
  {2016})}\BibitemShut {NoStop}%
\bibitem [{\citenamefont {Modak}\ and\ \citenamefont
  {Aravinda}(2022)}]{modak2022observational}%
  \BibitemOpen
  \bibfield  {author} {\bibinfo {author} {\bibfnamefont {Ranjan}\ \bibnamefont
  {Modak}}\ and\ \bibinfo {author} {\bibfnamefont {S}~\bibnamefont
  {Aravinda}},\ }\bibfield  {title} {\enquote {\bibinfo {title} {Observational
  entropic study of anderson localization},}\ }\href@noop {} {\bibfield
  {journal} {\bibinfo  {journal} {arXiv preprint arXiv:2209.10273}\ } (\bibinfo
  {year} {2022})}\BibitemShut {NoStop}%
\bibitem [{\citenamefont {Haake}\ \emph {et~al.}(1987)\citenamefont {Haake},
  \citenamefont {Ku{\'s}},\ and\ \citenamefont {Scharf}}]{haake1987classical}%
  \BibitemOpen
  \bibfield  {author} {\bibinfo {author} {\bibfnamefont {Fritz}\ \bibnamefont
  {Haake}}, \bibinfo {author} {\bibfnamefont {M}~\bibnamefont {Ku{\'s}}}, \
  and\ \bibinfo {author} {\bibfnamefont {Rainer}\ \bibnamefont {Scharf}},\
  }\bibfield  {title} {\enquote {\bibinfo {title} {Classical and quantum chaos
  for a kicked top},}\ }\href@noop {} {\bibfield  {journal} {\bibinfo
  {journal} {Zeitschrift f{\"u}r Physik B Condensed Matter}\ }\textbf {\bibinfo
  {volume} {65}},\ \bibinfo {pages} {381--395} (\bibinfo {year}
  {1987})}\BibitemShut {NoStop}%
\bibitem [{\citenamefont {Peres}(1997)}]{peres1997quantum}%
  \BibitemOpen
  \bibfield  {author} {\bibinfo {author} {\bibfnamefont {Asher}\ \bibnamefont
  {Peres}},\ }\href@noop {} {\emph {\bibinfo {title} {Quantum theory: concepts
  and methods}}}\ (\bibinfo  {publisher} {Springer},\ \bibinfo {year}
  {1997})\BibitemShut {NoStop}%
\bibitem [{\citenamefont {Chaudhury}\ \emph {et~al.}(2009)\citenamefont
  {Chaudhury}, \citenamefont {Smith}, \citenamefont {Anderson}, \citenamefont
  {Ghose},\ and\ \citenamefont {Jessen}}]{chaudhury2009quantum}%
  \BibitemOpen
  \bibfield  {author} {\bibinfo {author} {\bibfnamefont {S}~\bibnamefont
  {Chaudhury}}, \bibinfo {author} {\bibfnamefont {A}~\bibnamefont {Smith}},
  \bibinfo {author} {\bibfnamefont {BE}~\bibnamefont {Anderson}}, \bibinfo
  {author} {\bibfnamefont {S}~\bibnamefont {Ghose}}, \ and\ \bibinfo {author}
  {\bibfnamefont {Poul~S}\ \bibnamefont {Jessen}},\ }\bibfield  {title}
  {\enquote {\bibinfo {title} {Quantum signatures of chaos in a kicked top},}\
  }\href@noop {} {\bibfield  {journal} {\bibinfo  {journal} {Nature}\ }\textbf
  {\bibinfo {volume} {461}},\ \bibinfo {pages} {768--771} (\bibinfo {year}
  {2009})}\BibitemShut {NoStop}%
\bibitem [{\citenamefont {Neill}\ \emph {et~al.}(2016)\citenamefont {Neill},
  \citenamefont {Roushan}, \citenamefont {Fang}, \citenamefont {Chen},
  \citenamefont {Kolodrubetz}, \citenamefont {Chen}, \citenamefont {Megrant},
  \citenamefont {Barends}, \citenamefont {Campbell}, \citenamefont {Chiaro}
  \emph {et~al.}}]{neill2016ergodic}%
  \BibitemOpen
  \bibfield  {author} {\bibinfo {author} {\bibfnamefont {Charles}\ \bibnamefont
  {Neill}}, \bibinfo {author} {\bibfnamefont {P}~\bibnamefont {Roushan}},
  \bibinfo {author} {\bibfnamefont {M}~\bibnamefont {Fang}}, \bibinfo {author}
  {\bibfnamefont {Y}~\bibnamefont {Chen}}, \bibinfo {author} {\bibfnamefont
  {M}~\bibnamefont {Kolodrubetz}}, \bibinfo {author} {\bibfnamefont
  {Z}~\bibnamefont {Chen}}, \bibinfo {author} {\bibfnamefont {A}~\bibnamefont
  {Megrant}}, \bibinfo {author} {\bibfnamefont {R}~\bibnamefont {Barends}},
  \bibinfo {author} {\bibfnamefont {B}~\bibnamefont {Campbell}}, \bibinfo
  {author} {\bibfnamefont {B}~\bibnamefont {Chiaro}},  \emph {et~al.},\
  }\bibfield  {title} {\enquote {\bibinfo {title} {Ergodic dynamics and
  thermalization in an isolated quantum system},}\ }\href@noop {} {\bibfield
  {journal} {\bibinfo  {journal} {Nature Physics}\ }\textbf {\bibinfo {volume}
  {12}},\ \bibinfo {pages} {1037--1041} (\bibinfo {year} {2016})}\BibitemShut
  {NoStop}%
\bibitem [{\citenamefont {Dogra}\ \emph {et~al.}(2019)\citenamefont {Dogra},
  \citenamefont {Madhok},\ and\ \citenamefont
  {Lakshminarayan}}]{dogra2019quantum}%
  \BibitemOpen
  \bibfield  {author} {\bibinfo {author} {\bibfnamefont {Shruti}\ \bibnamefont
  {Dogra}}, \bibinfo {author} {\bibfnamefont {Vaibhav}\ \bibnamefont {Madhok}},
  \ and\ \bibinfo {author} {\bibfnamefont {Arul}\ \bibnamefont
  {Lakshminarayan}},\ }\bibfield  {title} {\enquote {\bibinfo {title} {Quantum
  signatures of chaos, thermalization, and tunneling in the exactly solvable
  few-body kicked top},}\ }\href@noop {} {\bibfield  {journal} {\bibinfo
  {journal} {Physical Review E}\ }\textbf {\bibinfo {volume} {99}},\ \bibinfo
  {pages} {062217} (\bibinfo {year} {2019})}\BibitemShut {NoStop}%
\bibitem [{\citenamefont {Wang}\ \emph {et~al.}(2004)\citenamefont {Wang},
  \citenamefont {Ghose}, \citenamefont {Sanders},\ and\ \citenamefont
  {Hu}}]{wang2004entanglement}%
  \BibitemOpen
  \bibfield  {author} {\bibinfo {author} {\bibfnamefont {Xiaoguang}\
  \bibnamefont {Wang}}, \bibinfo {author} {\bibfnamefont {Shohini}\
  \bibnamefont {Ghose}}, \bibinfo {author} {\bibfnamefont {Barry~C}\
  \bibnamefont {Sanders}}, \ and\ \bibinfo {author} {\bibfnamefont {Bambi}\
  \bibnamefont {Hu}},\ }\bibfield  {title} {\enquote {\bibinfo {title}
  {Entanglement as a signature of quantum chaos},}\ }\href@noop {} {\bibfield
  {journal} {\bibinfo  {journal} {Physical Review E}\ }\textbf {\bibinfo
  {volume} {70}},\ \bibinfo {pages} {016217} (\bibinfo {year}
  {2004})}\BibitemShut {NoStop}%
\bibitem [{\citenamefont {Stamatiou}\ and\ \citenamefont
  {Ghikas}(2007)}]{stamatiou2007quantum}%
  \BibitemOpen
  \bibfield  {author} {\bibinfo {author} {\bibfnamefont {George}\ \bibnamefont
  {Stamatiou}}\ and\ \bibinfo {author} {\bibfnamefont {Demetris~PK}\
  \bibnamefont {Ghikas}},\ }\bibfield  {title} {\enquote {\bibinfo {title}
  {Quantum entanglement dependence on bifurcations and scars in non-autonomous
  systems. the case of quantum kicked top},}\ }\href@noop {} {\bibfield
  {journal} {\bibinfo  {journal} {Physics Letters A}\ }\textbf {\bibinfo
  {volume} {368}},\ \bibinfo {pages} {206--214} (\bibinfo {year}
  {2007})}\BibitemShut {NoStop}%
\bibitem [{\citenamefont {Ghose}\ \emph {et~al.}(2008)\citenamefont {Ghose},
  \citenamefont {Stock}, \citenamefont {Jessen}, \citenamefont {Lal},\ and\
  \citenamefont {Silberfarb}}]{ghose2008chaos}%
  \BibitemOpen
  \bibfield  {author} {\bibinfo {author} {\bibfnamefont {Shohini}\ \bibnamefont
  {Ghose}}, \bibinfo {author} {\bibfnamefont {Rene}\ \bibnamefont {Stock}},
  \bibinfo {author} {\bibfnamefont {Poul}\ \bibnamefont {Jessen}}, \bibinfo
  {author} {\bibfnamefont {Roshan}\ \bibnamefont {Lal}}, \ and\ \bibinfo
  {author} {\bibfnamefont {Andrew}\ \bibnamefont {Silberfarb}},\ }\bibfield
  {title} {\enquote {\bibinfo {title} {Chaos, entanglement, and decoherence in
  the quantum kicked top},}\ }\href@noop {} {\bibfield  {journal} {\bibinfo
  {journal} {Physical Review A}\ }\textbf {\bibinfo {volume} {78}},\ \bibinfo
  {pages} {042318} (\bibinfo {year} {2008})}\BibitemShut {NoStop}%
\bibitem [{\citenamefont {Prosen}\ \emph {et~al.}(2003)\citenamefont {Prosen},
  \citenamefont {Seligman},\ and\ \citenamefont
  {{\v{Z}}nidari{\v{c}}}}]{prosen2003theory}%
  \BibitemOpen
  \bibfield  {author} {\bibinfo {author} {\bibfnamefont {Toma{\v{z}}}\
  \bibnamefont {Prosen}}, \bibinfo {author} {\bibfnamefont {Thomas~H}\
  \bibnamefont {Seligman}}, \ and\ \bibinfo {author} {\bibfnamefont {Marko}\
  \bibnamefont {{\v{Z}}nidari{\v{c}}}},\ }\bibfield  {title} {\enquote
  {\bibinfo {title} {Theory of quantum loschmidt echoes},}\ }\href@noop {}
  {\bibfield  {journal} {\bibinfo  {journal} {Progress of Theoretical Physics
  Supplement}\ }\textbf {\bibinfo {volume} {150}},\ \bibinfo {pages} {200--228}
  (\bibinfo {year} {2003})}\BibitemShut {NoStop}%
\bibitem [{\citenamefont {Schack}\ \emph {et~al.}(1994)\citenamefont {Schack},
  \citenamefont {D’Ariano},\ and\ \citenamefont
  {Caves}}]{schack1994hypersensitivity}%
  \BibitemOpen
  \bibfield  {author} {\bibinfo {author} {\bibfnamefont {R{\"u}diger}\
  \bibnamefont {Schack}}, \bibinfo {author} {\bibfnamefont {Giacomo~M}\
  \bibnamefont {D’Ariano}}, \ and\ \bibinfo {author} {\bibfnamefont
  {Carlton~M}\ \bibnamefont {Caves}},\ }\bibfield  {title} {\enquote {\bibinfo
  {title} {Hypersensitivity to perturbation in the quantum kicked top},}\
  }\href@noop {} {\bibfield  {journal} {\bibinfo  {journal} {Physical Review
  E}\ }\textbf {\bibinfo {volume} {50}},\ \bibinfo {pages} {972} (\bibinfo
  {year} {1994})}\BibitemShut {NoStop}%
\bibitem [{\citenamefont {Sahu}\ \emph {et~al.}(2022)\citenamefont {Sahu},
  \citenamefont {Sreeram},\ and\ \citenamefont {Madhok}}]{sahu2022effect}%
  \BibitemOpen
  \bibfield  {author} {\bibinfo {author} {\bibfnamefont {Abinash}\ \bibnamefont
  {Sahu}}, \bibinfo {author} {\bibfnamefont {PG}~\bibnamefont {Sreeram}}, \
  and\ \bibinfo {author} {\bibfnamefont {Vaibhav}\ \bibnamefont {Madhok}},\
  }\bibfield  {title} {\enquote {\bibinfo {title} {Effect of chaos on
  information gain in quantum tomography},}\ }\href@noop {} {\bibfield
  {journal} {\bibinfo  {journal} {Physical Review E}\ }\textbf {\bibinfo
  {volume} {106}},\ \bibinfo {pages} {024209} (\bibinfo {year}
  {2022})}\BibitemShut {NoStop}%
\bibitem [{\citenamefont {{\v{S}}afr{\'a}nek}\ and\ \citenamefont
  {Thingna}(2020)}]{safranek_gen}%
  \BibitemOpen
  \bibfield  {author} {\bibinfo {author} {\bibfnamefont {Dominik}\ \bibnamefont
  {{\v{S}}afr{\'a}nek}}\ and\ \bibinfo {author} {\bibfnamefont {Juzar}\
  \bibnamefont {Thingna}},\ }\href {\doibase 10.48550/ARXIV.2007.07246}
  {\enquote {\bibinfo {title} {Quantifying information extraction using
  generalized quantum measurements},}\ } (\bibinfo {year} {2020})\BibitemShut
  {NoStop}%
\bibitem [{\citenamefont {Buscemi}\ \emph
  {et~al.}(2022{\natexlab{b}})\citenamefont {Buscemi}, \citenamefont
  {Schindler},\ and\ \citenamefont
  {{\v{S}}afr{\'a}nek}}]{buscemi2022observational}%
  \BibitemOpen
  \bibfield  {author} {\bibinfo {author} {\bibfnamefont {Francesco}\
  \bibnamefont {Buscemi}}, \bibinfo {author} {\bibfnamefont {Joseph}\
  \bibnamefont {Schindler}}, \ and\ \bibinfo {author} {\bibfnamefont {Dominik}\
  \bibnamefont {{\v{S}}afr{\'a}nek}},\ }\bibfield  {title} {\enquote {\bibinfo
  {title} {Observational entropy, coarse quantum states, and petz recovery:
  information-theoretic properties and bounds},}\ }\href@noop {} {\bibfield
  {journal} {\bibinfo  {journal} {arXiv:2209.03803v1}\ } (\bibinfo {year}
  {2022}{\natexlab{b}})}\BibitemShut {NoStop}%
\bibitem [{\citenamefont {Buscemi}(2022)}]{Busc_talk}%
  \BibitemOpen
  \bibfield  {author} {\bibinfo {author} {\bibfnamefont {Francesco}\
  \bibnamefont {Buscemi}},\ }\href
  {https://www.youtube.com/watch?v=YJuC-hbSvxs} {\enquote {\bibinfo {title}
  {Observational entropy, coarse-grained states, and irretrodictability},}\ }
  (\bibinfo {year} {2022}),\ \bibinfo {note} {thirds Kyoto Workshop on Quantum
  Information, Computation, and Foundations}\BibitemShut {NoStop}%
\bibitem [{\citenamefont {Buscemi}\ \emph {et~al.}(2023)\citenamefont
  {Buscemi}, \citenamefont {{\v{S}}afr{\'a}nek}, \citenamefont {Schindler},\
  and\ \citenamefont {Scarani}}]{Busc_retro}%
  \BibitemOpen
  \bibfield  {author} {\bibinfo {author} {\bibfnamefont {Francesco}\
  \bibnamefont {Buscemi}}, \bibinfo {author} {\bibfnamefont {Dominik}\
  \bibnamefont {{\v{S}}afr{\'a}nek}}, \bibinfo {author} {\bibfnamefont
  {Joseph}\ \bibnamefont {Schindler}}, \ and\ \bibinfo {author} {\bibfnamefont
  {Valerio}\ \bibnamefont {Scarani}},\ }\href@noop {} {} (\bibinfo {year}
  {2023}),\ \bibinfo {note} {manuscript under preparation}\BibitemShut
  {NoStop}%
\bibitem [{\citenamefont {Buscemi}\ and\ \citenamefont
  {Scarani}(2021)}]{buscemi2021fluctuation}%
  \BibitemOpen
  \bibfield  {author} {\bibinfo {author} {\bibfnamefont {Francesco}\
  \bibnamefont {Buscemi}}\ and\ \bibinfo {author} {\bibfnamefont {Valerio}\
  \bibnamefont {Scarani}},\ }\bibfield  {title} {\enquote {\bibinfo {title}
  {Fluctuation theorems from bayesian retrodiction},}\ }\href@noop {}
  {\bibfield  {journal} {\bibinfo  {journal} {Physical Review E}\ }\textbf
  {\bibinfo {volume} {103}},\ \bibinfo {pages} {052111} (\bibinfo {year}
  {2021})}\BibitemShut {NoStop}%
\bibitem [{\citenamefont {Aw}\ \emph {et~al.}(2021)\citenamefont {Aw},
  \citenamefont {Buscemi},\ and\ \citenamefont {Scarani}}]{aw2021fluctuation}%
  \BibitemOpen
  \bibfield  {author} {\bibinfo {author} {\bibfnamefont {Clive~Cenxin}\
  \bibnamefont {Aw}}, \bibinfo {author} {\bibfnamefont {Francesco}\
  \bibnamefont {Buscemi}}, \ and\ \bibinfo {author} {\bibfnamefont {Valerio}\
  \bibnamefont {Scarani}},\ }\bibfield  {title} {\enquote {\bibinfo {title}
  {Fluctuation theorems with retrodiction rather than reverse processes},}\
  }\href@noop {} {\bibfield  {journal} {\bibinfo  {journal} {AVS Quantum
  Science}\ }\textbf {\bibinfo {volume} {3}},\ \bibinfo {pages} {045601}
  (\bibinfo {year} {2021})}\BibitemShut {NoStop}%
\bibitem [{\citenamefont {Kullback}\ and\ \citenamefont
  {Leibler}(1951)}]{kullback1951information}%
  \BibitemOpen
  \bibfield  {author} {\bibinfo {author} {\bibfnamefont {Solomon}\ \bibnamefont
  {Kullback}}\ and\ \bibinfo {author} {\bibfnamefont {Richard~A}\ \bibnamefont
  {Leibler}},\ }\bibfield  {title} {\enquote {\bibinfo {title} {On information
  and sufficiency},}\ }\href@noop {} {\bibfield  {journal} {\bibinfo  {journal}
  {The annals of mathematical statistics}\ }\textbf {\bibinfo {volume} {22}},\
  \bibinfo {pages} {79--86} (\bibinfo {year} {1951})}\BibitemShut {NoStop}%
\bibitem [{\citenamefont {Cover}(1999)}]{cover1999elements}%
  \BibitemOpen
  \bibfield  {author} {\bibinfo {author} {\bibfnamefont {Thomas~M}\
  \bibnamefont {Cover}},\ }\href@noop {} {\emph {\bibinfo {title} {Elements of
  information theory}}}\ (\bibinfo  {publisher} {John Wiley \& Sons},\ \bibinfo
  {year} {1999})\BibitemShut {NoStop}%
\bibitem [{\citenamefont {Umegaki}(1962)}]{umegaki1962conditional}%
  \BibitemOpen
  \bibfield  {author} {\bibinfo {author} {\bibfnamefont {Hisaharu}\
  \bibnamefont {Umegaki}},\ }\bibfield  {title} {\enquote {\bibinfo {title}
  {Conditional expectation in an operator algebra, iv (entropy and
  information)},}\ }in\ \href@noop {} {\emph {\bibinfo {booktitle} {Kodai
  Mathematical Seminar Reports}}},\ Vol.~\bibinfo {volume} {14}\ (\bibinfo
  {organization} {Department of Mathematics, Tokyo Institute of Technology},\
  \bibinfo {year} {1962})\ pp.\ \bibinfo {pages} {59--85}\BibitemShut {NoStop}%
\bibitem [{\citenamefont {Wilde}(2013)}]{wilde2013quantum}%
  \BibitemOpen
  \bibfield  {author} {\bibinfo {author} {\bibfnamefont {Mark~M}\ \bibnamefont
  {Wilde}},\ }\href@noop {} {\emph {\bibinfo {title} {Quantum information
  theory}}}\ (\bibinfo  {publisher} {Cambridge University Press},\ \bibinfo
  {year} {2013})\BibitemShut {NoStop}%
\bibitem [{\citenamefont {Asplund}\ and\ \citenamefont
  {Berenstein}(2016)}]{asplund2016entanglement}%
  \BibitemOpen
  \bibfield  {author} {\bibinfo {author} {\bibfnamefont {Curtis~T}\
  \bibnamefont {Asplund}}\ and\ \bibinfo {author} {\bibfnamefont {David}\
  \bibnamefont {Berenstein}},\ }\bibfield  {title} {\enquote {\bibinfo {title}
  {Entanglement entropy converges to classical entropy around periodic
  orbits},}\ }\href@noop {} {\bibfield  {journal} {\bibinfo  {journal} {Annals
  of Physics}\ }\textbf {\bibinfo {volume} {366}},\ \bibinfo {pages} {113--132}
  (\bibinfo {year} {2016})}\BibitemShut {NoStop}%
\bibitem [{\citenamefont {Bianchi}\ \emph {et~al.}(2018)\citenamefont
  {Bianchi}, \citenamefont {Hackl},\ and\ \citenamefont
  {Yokomizo}}]{bianchi2018linear}%
  \BibitemOpen
  \bibfield  {author} {\bibinfo {author} {\bibfnamefont {Eugenio}\ \bibnamefont
  {Bianchi}}, \bibinfo {author} {\bibfnamefont {Lucas}\ \bibnamefont {Hackl}},
  \ and\ \bibinfo {author} {\bibfnamefont {Nelson}\ \bibnamefont {Yokomizo}},\
  }\bibfield  {title} {\enquote {\bibinfo {title} {Linear growth of the
  entanglement entropy and the kolmogorov-sinai rate},}\ }\href@noop {}
  {\bibfield  {journal} {\bibinfo  {journal} {Journal of High Energy Physics}\
  }\textbf {\bibinfo {volume} {2018}},\ \bibinfo {pages} {1--70} (\bibinfo
  {year} {2018})}\BibitemShut {NoStop}%
\bibitem [{\citenamefont {Lerose}\ and\ \citenamefont
  {Pappalardi}(2020)}]{lerose2020bridging}%
  \BibitemOpen
  \bibfield  {author} {\bibinfo {author} {\bibfnamefont {Alessio}\ \bibnamefont
  {Lerose}}\ and\ \bibinfo {author} {\bibfnamefont {Silvia}\ \bibnamefont
  {Pappalardi}},\ }\bibfield  {title} {\enquote {\bibinfo {title} {Bridging
  entanglement dynamics and chaos in semiclassical systems},}\ }\href@noop {}
  {\bibfield  {journal} {\bibinfo  {journal} {Physical Review A}\ }\textbf
  {\bibinfo {volume} {102}},\ \bibinfo {pages} {032404} (\bibinfo {year}
  {2020})}\BibitemShut {NoStop}%
\bibitem [{\citenamefont {Modak}\ \emph {et~al.}(2020)\citenamefont {Modak},
  \citenamefont {Alba},\ and\ \citenamefont
  {Calabrese}}]{modak2020entanglement}%
  \BibitemOpen
  \bibfield  {author} {\bibinfo {author} {\bibfnamefont {Ranjan}\ \bibnamefont
  {Modak}}, \bibinfo {author} {\bibfnamefont {Vincenzo}\ \bibnamefont {Alba}},
  \ and\ \bibinfo {author} {\bibfnamefont {Pasquale}\ \bibnamefont
  {Calabrese}},\ }\bibfield  {title} {\enquote {\bibinfo {title} {Entanglement
  revivals as a probe of scrambling in finite quantum systems},}\ }\href@noop
  {} {\bibfield  {journal} {\bibinfo  {journal} {Journal of Statistical
  Mechanics: Theory and Experiment}\ }\textbf {\bibinfo {volume} {2020}},\
  \bibinfo {pages} {083110} (\bibinfo {year} {2020})}\BibitemShut {NoStop}%
\bibitem [{\citenamefont {Peres}(1984)}]{peres1984stability}%
  \BibitemOpen
  \bibfield  {author} {\bibinfo {author} {\bibfnamefont {Asher}\ \bibnamefont
  {Peres}},\ }\bibfield  {title} {\enquote {\bibinfo {title} {Stability of
  quantum motion in chaotic and regular systems},}\ }\href@noop {} {\bibfield
  {journal} {\bibinfo  {journal} {Physical Review A}\ }\textbf {\bibinfo
  {volume} {30}},\ \bibinfo {pages} {1610} (\bibinfo {year}
  {1984})}\BibitemShut {NoStop}%
\bibitem [{\citenamefont {Jalabert}\ and\ \citenamefont
  {Pastawski}(2001)}]{jalabert2001environment}%
  \BibitemOpen
  \bibfield  {author} {\bibinfo {author} {\bibfnamefont {Rodolfo~A}\
  \bibnamefont {Jalabert}}\ and\ \bibinfo {author} {\bibfnamefont {Horacio~M}\
  \bibnamefont {Pastawski}},\ }\bibfield  {title} {\enquote {\bibinfo {title}
  {Environment-independent decoherence rate in classically chaotic systems},}\
  }\href@noop {} {\bibfield  {journal} {\bibinfo  {journal} {Physical review
  letters}\ }\textbf {\bibinfo {volume} {86}},\ \bibinfo {pages} {2490}
  (\bibinfo {year} {2001})}\BibitemShut {NoStop}%
\bibitem [{\citenamefont {Xu}\ \emph {et~al.}(2020)\citenamefont {Xu},
  \citenamefont {Scaffidi},\ and\ \citenamefont {Cao}}]{xu2020does}%
  \BibitemOpen
  \bibfield  {author} {\bibinfo {author} {\bibfnamefont {Tianrui}\ \bibnamefont
  {Xu}}, \bibinfo {author} {\bibfnamefont {Thomas}\ \bibnamefont {Scaffidi}}, \
  and\ \bibinfo {author} {\bibfnamefont {Xiangyu}\ \bibnamefont {Cao}},\
  }\bibfield  {title} {\enquote {\bibinfo {title} {Does scrambling equal
  chaos?}}\ }\href@noop {} {\bibfield  {journal} {\bibinfo  {journal} {Physical
  review letters}\ }\textbf {\bibinfo {volume} {124}},\ \bibinfo {pages}
  {140602} (\bibinfo {year} {2020})}\BibitemShut {NoStop}%
\bibitem [{\citenamefont {Kidd}\ \emph {et~al.}(2021)\citenamefont {Kidd},
  \citenamefont {Safavi-Naini},\ and\ \citenamefont {Corney}}]{kidd2021saddle}%
  \BibitemOpen
  \bibfield  {author} {\bibinfo {author} {\bibfnamefont {RA}~\bibnamefont
  {Kidd}}, \bibinfo {author} {\bibfnamefont {A}~\bibnamefont {Safavi-Naini}}, \
  and\ \bibinfo {author} {\bibfnamefont {JF}~\bibnamefont {Corney}},\
  }\bibfield  {title} {\enquote {\bibinfo {title} {Saddle-point scrambling
  without thermalization},}\ }\href@noop {} {\bibfield  {journal} {\bibinfo
  {journal} {Physical Review A}\ }\textbf {\bibinfo {volume} {103}},\ \bibinfo
  {pages} {033304} (\bibinfo {year} {2021})}\BibitemShut {NoStop}%
\bibitem [{\citenamefont {G{\"a}rttner}\ \emph {et~al.}(2017)\citenamefont
  {G{\"a}rttner}, \citenamefont {Bohnet}, \citenamefont {Safavi-Naini},
  \citenamefont {Wall}, \citenamefont {Bollinger},\ and\ \citenamefont
  {Rey}}]{garttner2017measuring}%
  \BibitemOpen
  \bibfield  {author} {\bibinfo {author} {\bibfnamefont {Martin}\ \bibnamefont
  {G{\"a}rttner}}, \bibinfo {author} {\bibfnamefont {Justin~G}\ \bibnamefont
  {Bohnet}}, \bibinfo {author} {\bibfnamefont {Arghavan}\ \bibnamefont
  {Safavi-Naini}}, \bibinfo {author} {\bibfnamefont {Michael~L}\ \bibnamefont
  {Wall}}, \bibinfo {author} {\bibfnamefont {John~J}\ \bibnamefont
  {Bollinger}}, \ and\ \bibinfo {author} {\bibfnamefont {Ana~Maria}\
  \bibnamefont {Rey}},\ }\bibfield  {title} {\enquote {\bibinfo {title}
  {Measuring out-of-time-order correlations and multiple quantum spectra in a
  trapped-ion quantum magnet},}\ }\href@noop {} {\bibfield  {journal} {\bibinfo
   {journal} {Nature Physics}\ }\textbf {\bibinfo {volume} {13}},\ \bibinfo
  {pages} {781--786} (\bibinfo {year} {2017})}\BibitemShut {NoStop}%
\bibitem [{\citenamefont {G{\"a}rttner}\ \emph {et~al.}(2018)\citenamefont
  {G{\"a}rttner}, \citenamefont {Hauke},\ and\ \citenamefont
  {Rey}}]{garttner2018relating}%
  \BibitemOpen
  \bibfield  {author} {\bibinfo {author} {\bibfnamefont {Martin}\ \bibnamefont
  {G{\"a}rttner}}, \bibinfo {author} {\bibfnamefont {Philipp}\ \bibnamefont
  {Hauke}}, \ and\ \bibinfo {author} {\bibfnamefont {Ana~Maria}\ \bibnamefont
  {Rey}},\ }\bibfield  {title} {\enquote {\bibinfo {title} {Relating
  out-of-time-order correlations to entanglement via multiple-quantum
  coherences},}\ }\href@noop {} {\bibfield  {journal} {\bibinfo  {journal}
  {Physical review letters}\ }\textbf {\bibinfo {volume} {120}},\ \bibinfo
  {pages} {040402} (\bibinfo {year} {2018})}\BibitemShut {NoStop}%
\bibitem [{\citenamefont {Modak}\ \emph {et~al.}(2014)\citenamefont {Modak},
  \citenamefont {Mukerjee},\ and\ \citenamefont {Ramaswamy}}]{modak.2014}%
  \BibitemOpen
  \bibfield  {author} {\bibinfo {author} {\bibfnamefont {Ranjan}\ \bibnamefont
  {Modak}}, \bibinfo {author} {\bibfnamefont {Subroto}\ \bibnamefont
  {Mukerjee}}, \ and\ \bibinfo {author} {\bibfnamefont {Sriram}\ \bibnamefont
  {Ramaswamy}},\ }\bibfield  {title} {\enquote {\bibinfo {title} {Universal
  power law in crossover from integrability to quantum chaos},}\ }\href
  {\doibase 10.1103/PhysRevB.90.075152} {\bibfield  {journal} {\bibinfo
  {journal} {Phys. Rev. B}\ }\textbf {\bibinfo {volume} {90}},\ \bibinfo
  {pages} {075152} (\bibinfo {year} {2014})}\BibitemShut {NoStop}%
\bibitem [{\citenamefont {D'Alessio}\ \emph {et~al.}(2016)\citenamefont
  {D'Alessio}, \citenamefont {Kafri}, \citenamefont {Polkovnikov},\ and\
  \citenamefont {Rigol}}]{d2016quantum}%
  \BibitemOpen
  \bibfield  {author} {\bibinfo {author} {\bibfnamefont {Luca}\ \bibnamefont
  {D'Alessio}}, \bibinfo {author} {\bibfnamefont {Yariv}\ \bibnamefont
  {Kafri}}, \bibinfo {author} {\bibfnamefont {Anatoli}\ \bibnamefont
  {Polkovnikov}}, \ and\ \bibinfo {author} {\bibfnamefont {Marcos}\
  \bibnamefont {Rigol}},\ }\bibfield  {title} {\enquote {\bibinfo {title} {From
  quantum chaos and eigenstate thermalization to statistical mechanics and
  thermodynamics},}\ }\href@noop {} {\bibfield  {journal} {\bibinfo  {journal}
  {Advances in Physics}\ }\textbf {\bibinfo {volume} {65}},\ \bibinfo {pages}
  {239--362} (\bibinfo {year} {2016})}\BibitemShut {NoStop}%
\bibitem [{\citenamefont {Goold}\ \emph
  {et~al.}(2016{\natexlab{b}})\citenamefont {Goold}, \citenamefont {Huber},
  \citenamefont {Riera}, \citenamefont {del Rio},\ and\ \citenamefont
  {Skrzypczyk}}]{GHR+16}%
  \BibitemOpen
  \bibfield  {author} {\bibinfo {author} {\bibfnamefont {John}\ \bibnamefont
  {Goold}}, \bibinfo {author} {\bibfnamefont {Marcus}\ \bibnamefont {Huber}},
  \bibinfo {author} {\bibfnamefont {Arnau}\ \bibnamefont {Riera}}, \bibinfo
  {author} {\bibfnamefont {Lídia}\ \bibnamefont {del Rio}}, \ and\ \bibinfo
  {author} {\bibfnamefont {Paul}\ \bibnamefont {Skrzypczyk}},\ }\bibfield
  {title} {\enquote {\bibinfo {title} {The role of quantum information in
  thermodynamics—a topical review},}\ }\href
  {http://stacks.iop.org/1751-8121/49/i=14/a=143001} {\bibfield  {journal}
  {\bibinfo  {journal} {Journal of Physics A: Mathematical and Theoretical}\
  }\textbf {\bibinfo {volume} {49}},\ \bibinfo {pages} {143001} (\bibinfo
  {year} {2016}{\natexlab{b}})}\BibitemShut {NoStop}%
\bibitem [{\citenamefont {Yukalov}(2011)}]{yukalov2011equilibration}%
  \BibitemOpen
  \bibfield  {author} {\bibinfo {author} {\bibfnamefont {VI}~\bibnamefont
  {Yukalov}},\ }\bibfield  {title} {\enquote {\bibinfo {title} {Equilibration
  and thermalization in finite quantum systems},}\ }\href@noop {} {\bibfield
  {journal} {\bibinfo  {journal} {Laser Physics Letters}\ }\textbf {\bibinfo
  {volume} {8}},\ \bibinfo {pages} {485} (\bibinfo {year} {2011})}\BibitemShut
  {NoStop}%
\bibitem [{\citenamefont {Seifert}(2012)}]{seifert2012stochastic}%
  \BibitemOpen
  \bibfield  {author} {\bibinfo {author} {\bibfnamefont {Udo}\ \bibnamefont
  {Seifert}},\ }\bibfield  {title} {\enquote {\bibinfo {title} {Stochastic
  thermodynamics, fluctuation theorems and molecular machines},}\ }\href@noop
  {} {\bibfield  {journal} {\bibinfo  {journal} {Reports on progress in
  physics}\ }\textbf {\bibinfo {volume} {75}},\ \bibinfo {pages} {126001}
  (\bibinfo {year} {2012})}\BibitemShut {NoStop}%
\end{thebibliography}%
\end{document}